\documentclass[a4paper,11pt]{article}
\usepackage{rotating}
\usepackage{jheppub} 
\usepackage[utf8]{inputenc}
\usepackage{placeins} 
\usepackage{chngpage}
\usepackage{amsmath,amsfonts}
\usepackage{graphicx}
\usepackage{color}
\usepackage{xcolor}
\usepackage{relsize}
\usepackage{epstopdf}
\usepackage{hyperref}
\usepackage{mathrsfs}
\usepackage{ragged2e}
\usepackage{amssymb}
\usepackage{placeins}
\usepackage[normalem]{ ulem }
\usepackage{amsthm}
\usepackage{comment}
\usepackage{graphics}
\usepackage{caption}
\usepackage{subcaption}
\usepackage{verbatim}
\usepackage{arydshln}
\usepackage{cprotect}
\usepackage{diagbox}

\usepackage{tabularx,booktabs}
\usepackage{multicol}
\usepackage{array,multirow}
\usepackage{booktabs}
\usepackage{colortbl}
\usepackage{float}

\def\gsim{\raise0.3ex\hbox{$\;>$\kern-0.75em\raise-1.1ex\hbox{$\sim\;$}}}
\def\lsim{\raise0.3ex\hbox{$\;<$\kern-0.75em\raise-1.1ex\hbox{$\sim\;$}}}

\makeatletter
\gdef\@fpheader{}
\vspace*{0cm}
\makeatother

\begin{document}

\title{SMEFT goes dark: Dark Matter models for four-fermion operators}

\author[a]{Ricardo Cepedello,}
\author[b]{Fabian Esser,}
\author[b]{Martin Hirsch}
\author[b,c]{and Veronica Sanz}
\affiliation[a]{Institut f\"ur Theoretische Physik und Astrophysik, 
Universit\"{a}t W\"{u}rzburg, 97074 Würzburg, Germany}
\affiliation[b]{Instituto de F\'isica Corpuscular (IFIC), 
Universidad de Valencia-CSIC, E-46980 Valencia, Spain}
\affiliation[c]{Department of Physics and Astronomy, 
University of Sussex, Brighton BN1 9QH, UK}

\emailAdd{ricardo.cepedello@physik.uni-wuerzburg.de}
\emailAdd{esser@ific.uv.es}
\emailAdd{mahirsch@ific.uv.es}
\emailAdd{veronica.sanz@uv.es}

\date{\today}

\abstract{ 

We study ultra-violet completions for $d=6$ four-fermion operators in the
standard model effective field theory (SMEFT), focusing on models that
contain cold dark matter candidates. Via a diagrammatic method, we
generate systematically lists of possible UV completions, with the aim
of providing sets of models, which are complete under certain, well
specified assumptions. Within these lists of models we rediscover many
known DM models, as diverse as $R$-parity conserving supersymmetry or
the scotogenic neutrino mass model. Our lists, however, also contain
many new constructions, which have not been studied in the
literature so far. We also briefly discuss how our DM models could be 
constrained by reinterpretations of LHC searches and the prospects for HL-LHC and future lepton colliders.

}

             
\maketitle


\section{Introduction\label{sec:intro}}

The search for new resonances at the LHC has so far come up
empty-handed. Consequently, interest has been shifting in the past few
years to the study of effective field theories (EFTs) and, after the
discovery of the Higgs boson at the LHC
\cite{ATLAS:2012yve,CMS:2012qbp}, the most widely used EFT approach is
the SMEFT (``standard model EFT'') \cite{Weinberg:1979sa,
  Buchmuller:1985jz, Grzadkowski:2010es, Lehman:2014jma,
  Lehman:2015coa, Henning:2015alf, Gripaios:2018zrz, Criado:2019ugp,
  Murphy:2020rsh, Li:2020gnx, Li:2020xlh, Liao:2020jmn}.

EFTs parametrise new physics (NP) as a series of non-renormalisable 
operators:
\begin{equation}\label{eq:Ops}
{\cal L} = {\cal L}_{d=4} +\sum_k \frac{c_k}{\Lambda^{d-4}}{\cal O}_k ,
\end{equation}
where $\Lambda$ is the energy scale of NP and the sum includes terms
from $d=5,\cdots$ up to the dimension required by the precision of the
experiment under consideration. Low energy probes and the highly
precise LEP experiments provide very stringent constraints on a number
of SMEFT operators \cite{Falkowski:2015krw,Falkowski:2017pss,
  Falkowski:2020pma}, which push the scale of new physics into the
multi-TeV region (or even hundreds of TeV in case of lepton-flavour
violating observables \cite{Crivellin:2017rmk}), if $c_k$ is of order one.
 
Of course, putting $c_k=1$ is an unrealistic assumption for nearly
every ultra-violet completion that one can construct for the different
${\cal O}_k$'s (with the notable exception of new, strongly coupled
sectors, see e.g. the discussion in Ref. \cite{Jenkins:2013fya}). In
particular, the new physics generating the non-renormalisable
operators (NROs) in \eqref{eq:Ops} may be such that all NP operators
appear only at 1-loop level. In that case, the ``natural'' assumption
for $c_k$ changes to $c_k= 1/(16\pi^2)$. In this class of models all
constraints on $\Lambda$ are relaxed and direct searches at the LHC
may be competitive -- or even superior -- to indirect constraints.

In general, there are two different classes of new physics models, in which
some or all of the NROs appear at the 1-loop level at leading
order. These are: (i) accidentals and (ii) symmetry protected UV
models. We have discussed (i) for the special case of four-fermion
(4F) operators in \cite{Cepedello:2022pyx}. Class (ii) symmetry
protected models, on the other hand, can easily be motivated from the
fact that the standard model has no candidate to explain the observed
dark matter content of the universe.

Weakly interacting massive particles (WIMPs) have been discussed as
the prime candidates for the cold dark matter (DM) in the universe for
many years, for a recent review on the status of WIMPs see, for
example, \cite{Arcadi:2017kky}.  In order to explain the observed DM
abundance today, WIMPs must be stable particles, or at least have
life-times exceeding (by far) the age of the universe
\cite{Audren:2014bca}. Stable WIMPs, however, require a protecting
symmetry. The simplest possible case is a $Z_2$ under which the WIMP
(and, possibly, other BSM particles in a given model) are odd, while
the SM field content is even. Such a setup implies that all (odd) BSM
fields can couple only in pairs to SM fields. It is trivial to see
that all contributions of these BSM states to SMEFT operators are then
one-loop suppressed.  Thus, such DM models will be only weakly
constrained by indirect searches for NROs and one should expect that
direct searches at the high-energy frontier, together with direct DM
detection experiments \cite{XENON:2018voc,PandaX-4T:2021bab}, will
have the best expectations to discover them. This simple
  observation forms the basic motivation to study such 1-loop models
  in our current work.

In this paper we discuss a systematic construction of (WIMP) dark
matter model variants based on a diagrammatic method
\cite{Cepedello:2018rfh,Cepedello:2019zqf,Cepedello:2022pyx} for
four-fermion (4F) operators. The rest of this paper is organised as
follows. In section \ref{sec:boxes} we discuss the basics of the
diagrammatic method and the list of phenomenologically allowed WIMP DM
candidates that we consider. Section \ref{sect:models} presents our
results: In subsection \ref{subsect:counting_models} we count the
number of models we find for four-fermion operators and their model
overlap with other four-fermion and fermion-Higgs operators. Next, in
subsection \ref{subsect:patterns} we mention common patterns in the
models we find and list all new particles that are featured by the
models for specific DM candidates in subsection
\ref{subsect:new_particles}. Then, in subsection \ref{subsect:match},
we examine the matching for a variety of example models\footnote{Note
  that the model files and matching results for the model examples
  discussed in this paper can be found in the auxiliary files attached
  to this paper. }. In section \ref{sect:pheno} we analyse the
phenomenology of these models.  We close with a short discussion.


\section{One-loop dark matter models for SMEFT 4F operators\label{sec:boxes}}

In this section we will outline our method to construct UV completions
for 4-fermion SMEFT operators with dark matter candidates.  The
discussion follows in parts \cite{Cepedello:2022pyx}, where the same
methods were used to construct 1-loop models with ``exit
particles''.\footnote{``Exit particles'', as defined in
  \cite{Cepedello:2022pyx} are simply fields that can couple linearly
  to SM fields.} We will therefore be rather brief. The main
difference to our previous work is that, in order to ensure the DM
stability, we will assume that some of the BSM fields will be odd
under a $Z_2$ symmetry.
\footnote{The symmetry does not necessarily have to be a $Z_2$.  Any
  symmetry that forbids the DM candidate to couple linearly to SM
  fields will be sufficient. $Z_2$ is just the simplest example.}
This difference leads to some distinct patterns, different from those
for models with ``exit particles", as will be discussed in subsection
\ref{subsect:patterns}.

\subsection{Diagrammatic approach}
\label{subsect:diagrammatic_approach}

\begin{figure}[t]
\begin{center}
\includegraphics[width=\linewidth]{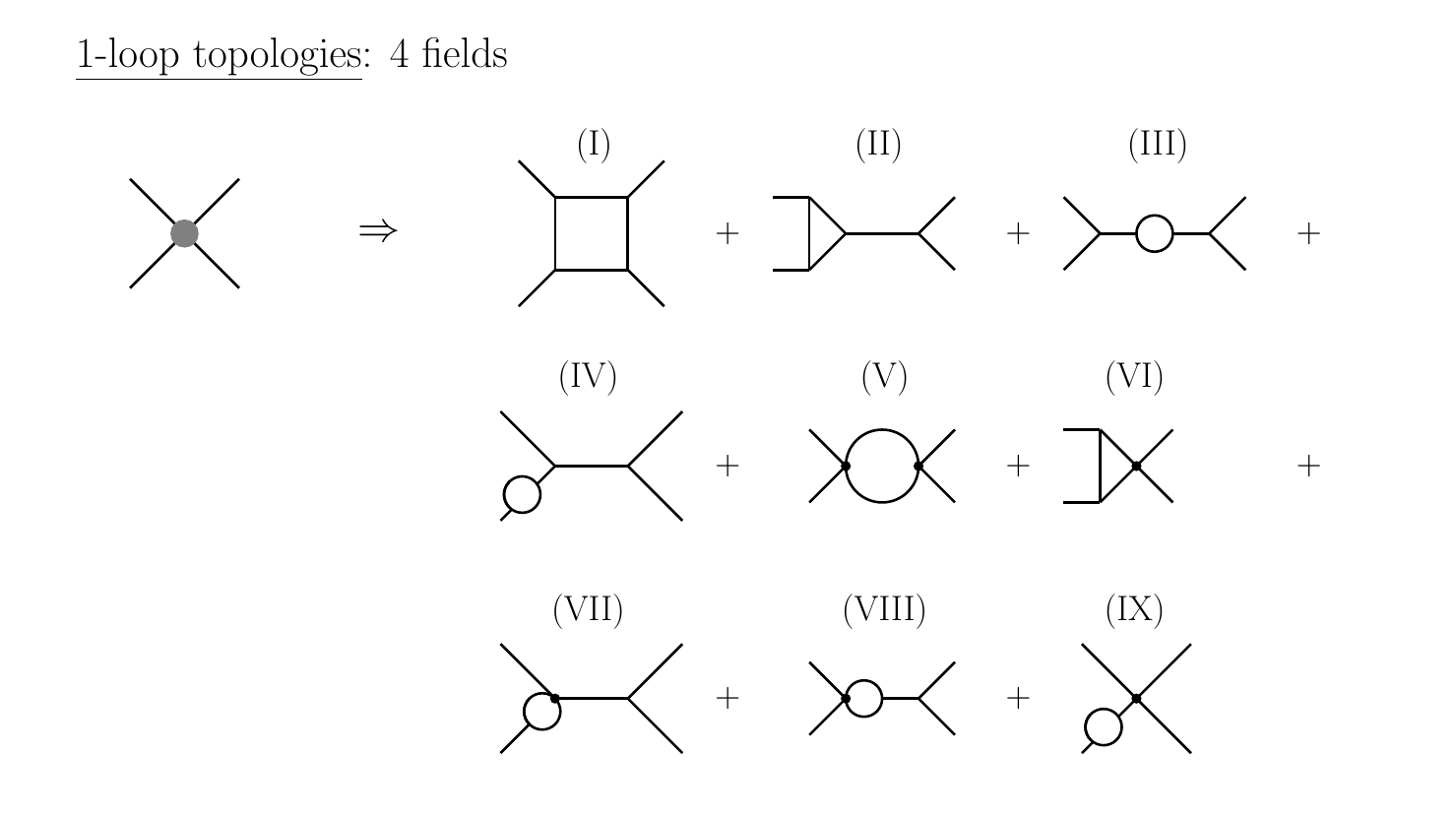}
\caption{There are in total nine 1-loop topologies for operators with
  four external fields, disregarding tadpole diagrams.  For discussion
  see text.}
\label{fig:1LPTopo}
\end{center}
\end{figure}

The procedure involves essentially three steps. First, for any given
SMEFT operator one can find all topologies that can generate the
operator at a chosen loop level. Here, topologies are just field lines
connected by certain types of vertices, without specifying the Lorentz
nature of the fields (scalars, fermions or vectors). Since we are
interested in renormalisable UV completions, only 3- and 4-point
vertices are allowed in this step. We do not consider topologies that
lead only to tadpole diagrams. This part of the calculation can be
easily automated, since topologies can be expressed in the form of
adjacency matrices. All 1-loop topologies for operators with four
external legs are shown in Fig.\ \ref{fig:1LPTopo}.

Specifying the outside particles to be fermions (4F operators),
only topologies I, II and III are left as interesting for DM model
constructions. Topologies V to IX contain at least one
4-point vertex connected to the external fermionic legs. These are
non-renormalisable vertices and, therefore, we discard them
immediately. Topology IV is also not interesting, since it leads only
to self-energy diagrams of an external fermion line.

\begin{figure}[t]
\begin{center}
\includegraphics[width=\linewidth]{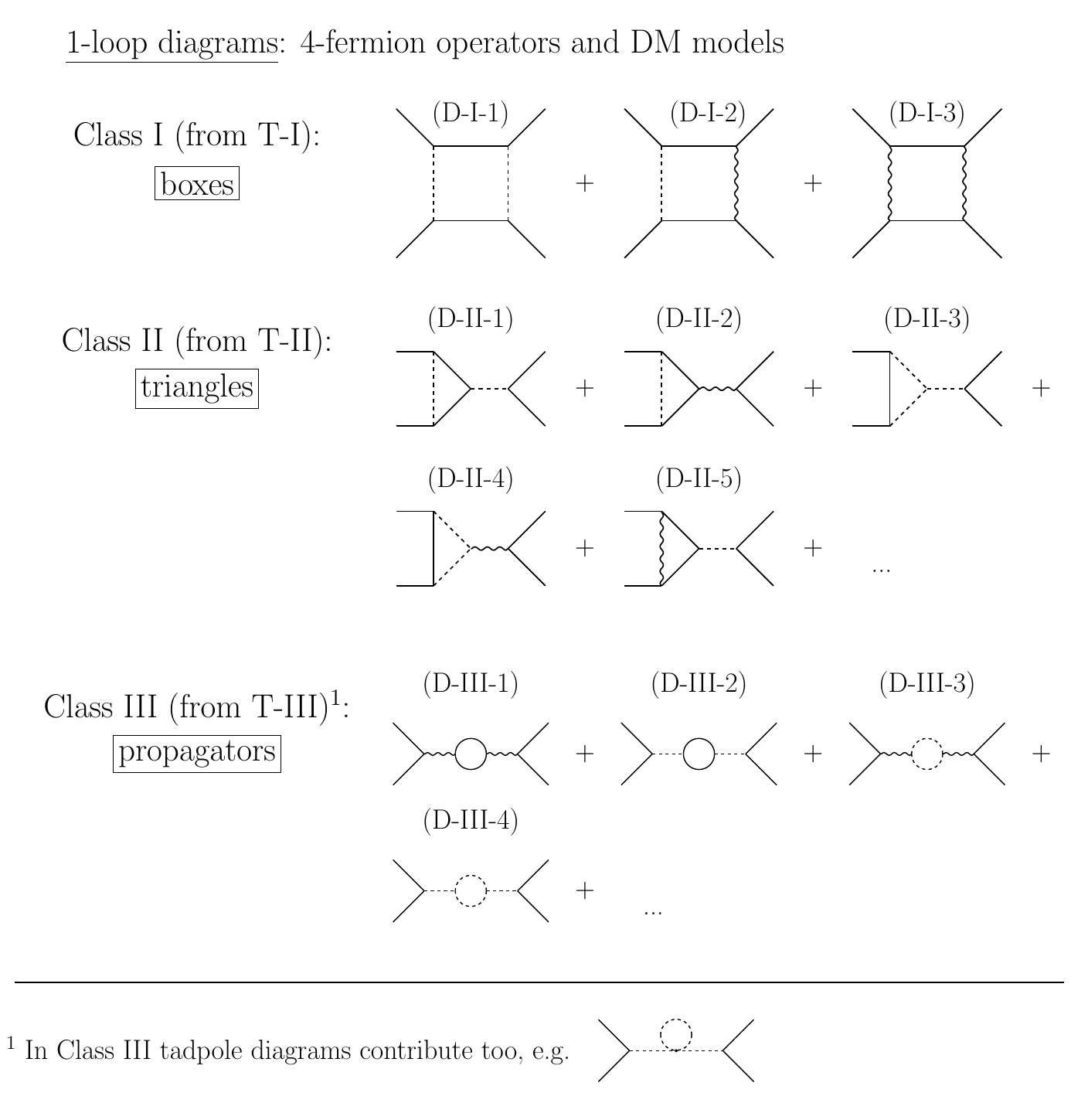}
\caption{1-loop diagrams for 4-fermion operators. There are 
three different classes of models: (i) Boxes (top), all BSM particles 
are odd; (ii) triangles (middle), in addition to the BSM fields 
that are odd, there is an even BSM field that acts as a portal. 
And, finally (iii) propagator corrections (bottom). The latter can 
yield  interesting models only for some special cases, see the 
discussion in the text.}
\label{fig:1LPDiags}
\end{center}
\end{figure}

The remaining three topologies can yield diagrams which lead to
phenomenologically interesting DM models. Fig.\ \ref{fig:1LPDiags}
shows a partial list. Class-I models come from box diagrams.
While there are, in principle, three possible diagrams, in the rest of
this paper we will concentrate on only diagrams with scalars and
fermions for simplicity, i.e. diagram D-I-1. Models with BSM vectors are, of course, also
possible, but the construction of fully consistent DM models with
vectors is much more intricate than for models with only new scalars,
see the discussion in \cite{Cepedello:2022pyx} and in particular
\cite{Fonseca:2016jbm,Fonseca:2022wtz}.

\begin{figure}[t]
\begin{center}
\includegraphics[width=0.5\linewidth]{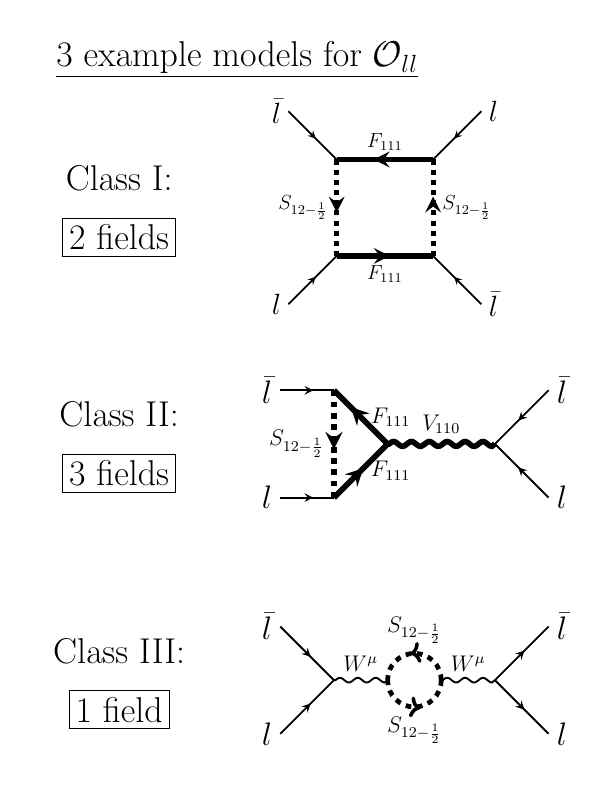}
\caption{The three simplest example models for class-I, class-II 
and class-III diagrams for one example 4-fermion operator, 
${\cal O}_{ll}$.}
\label{fig:OllMdls}
\end{center}
\end{figure}

DM models from box diagrams need at least two BSM fields. A very simple
example model for ${\cal O}_{ll}$ is shown in Fig.\ \ref{fig:OllMdls}.  
All particles inside the box need to be $Z_2$ odd in order to
guarantee that the lightest loop particle is stable. In this example
the neutral component of $S_{1,2,1/2}$ can be the DM candidate. The
model is a simple extension of the well-known ``inert doublet'' model
\cite{LopezHonorez:2006gr}.

Class-II models can be constructed from triangle diagrams. In
Fig.\ \ref{fig:1LPDiags} we show an incomplete list for this class of
diagrams. Again, triangles need at least two additional BSM fields
which are $Z_2$ odd.  Different from box diagrams, however, triangles
need one additional BSM field that couples linearly to SM
fields,\footnote{If the additional field connecting at tree-level to
  the fermions is itself a SM field, the DM loop is only a vertex
  correction for some tree-level SM coupling.}  an example model is
shown again in Fig.\ \ref{fig:OllMdls}. One can understand class-II
models as special realizations of ``portal'' DM models, where the
$Z_2$-even BSM particle connects the SM with the dark sector. While
these constructions can yield valid DM models, one can not expect to
obtain any meaningful constraints on such models from the study of 4F
operators. This can be easily understood: The additional $Z_2$-even
BSM field will generate the 4F operator under consideration at
tree-level. Thus, the DM loop is only a minor (and for all practical
purposes negligible) correction. We therefore will not discuss models
in class-II further in this paper. Note, however, that the particle
content of the loop with the coupling to $F_1$ and $F_2$ on the left
of the diagram, will give a DM model contributing to the box diagram
for a 4F operator containing $(F_1F_2)^2$ and thus the $Z_2$ odd
particle contents of triangle models appear in our list of box models.

Finally, in class-III we find models that give diagrams corresponding
to propagator corrections. Here, we have to distinguish three
sub-classes.  The first subclass (subclass III-a) are diagrams in
which a DM candidate, non-singlet under the SM group, can couple to a
SM gauge field. An example model is shown in Fig.\ \ref{fig:OllMdls}.
One can write down redundant $d=6$ operators of the type ${\cal
  R}_{2X} \propto (D_{\mu}X^{\mu\nu})^2$, where $X^{\mu\nu}$
  generically stands for a field strength tensor. Upon using the
equation of motion (EOM) for the field strength tensor, these types of
operators are expressed in terms of 4F operators with fermion pairs
coupling to $X^{\mu\nu}$. Thus, any BSM particle contributing to
${\cal R}_{2X}$ will contribute to 4F operators in the Warsaw basis.
Subclass III-a are valid DM models, but not particularly interesting
for us for the following reason. These constructions are models in
which the SM particle content is simply extended by the DM candidate
under consideration. Thus, giving the list of ``valid'' DM candidates
is equivalent of giving the list of DM models with {\em exactly one
  BSM field} and we have nothing new to add to this class from the
point of view of model-building.  Note that in this class of single
field models the matching coefficients of the different 4F operators
will be strictly related and always generation diagonal. For example,
a model with only one copy of $F_{1,3,0}$ would give that matching
conditions ${\cal O}_{ll}={\cal O}_{qq}^{3} =(1/2) {\cal O}_{lq}^{3}$,
while all other 4F operators get zero contribution. However, all
Wilson coefficients would be suppressed by $g_2^4$ and proportional to
rather small coefficients.

The second subclass, III-b, occurs only in some particular 4F
operators. In this subclass the internal particle coupling at
tree-level to a fermion pair is the SM Higgs field. Propagator
corrections of the type D-III-2 and D-III-4 can occur in DM models
with a DM candidate and one additional fermion (D-III-2) or scalar
(D-III-4). Such two field extensions of the SM will give contributions
to some 4F operators, but will be suppressed by SM Yukawa couplings. They are
therefore not interesting phenomenologically (except for operators
containing a pair of top quarks, which we do not discuss in this
paper) and we will not consider this class in further detail.

The remaining subclass, III-c, contains diagrams in which one (or
both) of the particles coupling to the outside fermions is a BSM
field. In this case, similar comments apply as to the triangle
diagrams: These contributions to the 4F operators will always be
generated in models with a ``portal'' field and some DM candidate, but
since the portal field must also couple to the SM fermions at
tree-level, the DM loop is a negligible correction to the 4F 
operator.

Thus, one can conclude that DM models that are phenomenologically
interesting for 4F operators are the ones found in box diagrams and we
will concentrate on discussing these in the rest of the paper. The
third and last step in the diagrammatic method then consists in
finding all valid particle insertions for this type of diagram for all
possible 4F operators. This is discussed next.

\subsection{Dark matter content}
\label{subsect:DM_content}

Every vertex of a box diagram connects one SM particle
with two undetermined BSM fields, meaning that the interaction is not fixed
unambiguously. This leads, in principle, to an infinite ``tower'' of
models that one could construct. However, in the current paper we are
interested only in models in which at least one particle in the loop
can be a good WIMP dark matter candidate.  This implies that, once the
list of possible DM candidates is known, the construction of all
possible models becomes a manageable combinatorical problem: One needs
to find all non-isomorphic permutations of the external fields in a given
operator and insert the chosen DM candidate in all possible
positions. Then, the other particles in the diagram are fixed and their
quantum numbers can be calculated. Finally, one eliminates all
duplicate models. This procedure is then repeated over all operators
and the full list of DM candidates one wishes to consider. 

Before discussing the models we still have to find the list of possible
DM WIMP candidates. Here, we consider as ``good'',
i.e.\ phenomenologically consistent, WIMP DM candidates only
colour-singlet, electrically neutral particles.\footnote{Thus, we will
  not discuss more exotic possibilities such as SIMPS (``strongly
  interacting massive particles'') or milli-charged dark matter and
  other dark sector models. For a review of these and other exotics,
  see for example \cite{Alexander:2016aln}.} The DM candidate should
also allow to fit correctly the observed DM relic density or, at
least, there should be regions in the parameter space of the model
where its relic abundance is not larger than $\Omega_{DM}$ as measured
by Planck \cite{Planck:2018vyg}. And, finally, it should also
obey various bounds from direct detection (DD) experiments and
other searches. The best DD limits are currently from experiments
using Xe \cite{XENON:2018voc,PandaX-4T:2021bab}.

What type of multiplets can be phenomenologically consistent DM WIMP
candidates? As far as we know, this question was originally addressed
in \cite{Cirelli:2005uq}. References 
\cite{Bottaro:2021snn,Bottaro:2022one} have recently provided a
detailed update. Further studies can be found, for example, in
\cite{Hambye:2009pw} and \cite{Belyaev:2022qnf}. The following
discussion draws heavily from the results of
\cite{Bottaro:2021snn,Bottaro:2022one}.

The annihilation cross section for the WIMP in the early universe
increases with the size of the $SU(2)$ representation.
References \cite{Bottaro:2021snn,Bottaro:2022one} showed that for $SU(2)$
multiplets larger than a ${\bf 13}$-plet the annihilation cross
section will violate the (s-wave) unitarity bound, thus limiting the
maximal size of $SU(2)$ representation allowed for DM candidates. This
upper limit is actually weaker than the criterion used originally in
\cite{Cirelli:2005uq}. The authors of \cite{Cirelli:2005uq} showed
that for any $SU(2)$ representation larger than quintuplet (fermions)
or ${\bf 7}$-plet (scalars), the running of $\alpha_2$ will hit a
Landau pole below the GUT (``grand unified theory'') scale. The list
of possible DM candidates given in \cite{Cirelli:2005uq} therefore
contains only multiplets up to a scalar ${\bf 7}$-plet.

We can roughly divide all the remaining multiplets into just two
cases: Models with DM candidates that are members of an $SU(2)$
multiplet with $Y=0$, and all others. The case $Y=0$ has been studied
in detail in \cite{Bottaro:2021snn}, $Y \ne 0$ has been treated in
\cite{Bottaro:2022one}.

The $Y=0$ case is the simpler one.  Here, the list consists of both
scalars and fermions that are odd $SU(2)$ multiplets,
i.e.\ representations $(1,n,0)$, with $n=1,3,5,\cdots$. For this case,
the neutral component of the multiplet has no tree-level coupling to
the $Z^0$ boson, thus the DD cross section is very suppressed and all
multiplets allowed by s-wave unitarity bounds survive current upper DD
limits \cite{Bottaro:2021snn}.\footnote{However, the future DARWIN experiment \cite{DARWIN:2016hyl} has the potential to rule out all non-trivial $Y=0$ multiplets as the main component of the DM, as long as we are not considering unnatural cancellations.}  In
section \ref{sect:models} we will discuss both models with fermion and
scalar candidates. The extension of these results to
the construction of models with larger multiplets is straightforward.

The $Y\ne 0$ case is more complicated, since DM candidates from
multiplets with $Y\ne 0$ are already excluded by DD limits
\cite{Cirelli:2005uq,Bottaro:2022one}, unless they fall within the
``inelastic dark matter'' class. Inelastic dark matter are DM
candidates with a sufficiently large mass splitting between the
CP-even and CP-odd components of a neutral state. Since the $Z^0$
boson is CP-odd, it will always couple ``off-diagonally'' between
the CP-even and CP-odd components. If the mass splitting
between these states is larger than the kinetic energy of the DM
particle, contributions from $Z^0$ to the DD cross sections are 
kinematically forbidden and, thus, $Y\ne 0$ DM candidates can survive 
the stringent DD bounds in this part of the parameter space.

From the model-building perspective, there is one important difference 
between scalars and fermions in this respect. For scalars with 
$Y=1/2$, one  can write down the following quartic coupling:
\begin{equation}\label{eq:pot12}
V \propto \lambda_5 \, \left[ S_{1,2n,1/2}^{\dagger} 
(T^a) S_{1,2n,1/2}^c \right] \left[ (H^c)^{\dagger}(\sigma^a/2) H \right] + \text{h.c.} \, .
\end{equation}
After electro-weak symmetry breaking the mass squared of the CP-even 
and CP-odd parts of the neutral component of the multiplet $S_{1,2n,1/2}$
receives contributions proportional to $\pm \lambda_5 v^2$, where
$v$ is the SM Higgs vacuum expectation value. Thus, the required 
mass splitting for inelastic dark matter can be generated from 
renormalisable terms of the model Lagrangian (for a sufficiently 
large value of $\lambda_5$). 

For fermions and for larger values of $Y$, however, this mass
splitting can be generated only via non-renormalisable operators and,
moreover, for $Y>1/2$ consistency requirements between the mass of the
DM and the energy scale of the NROs rule out all multiplets, except
$F_{1,3(5),1}$ and $S_{1,3(5),1}$, as inelastic DM \cite{Bottaro:2022one}. 
Since it is our aim to deconstruct the 4F operators into renormalizable 
models we do not consider models containing $Y\ne 0$ in detail, 
except for $S_{1,2n,1/2}$. Note, however, that our methods could be 
easily applied to these other multiplets as well, if one were interested 
only in reconstructing the particle content of the models.


\section{Model candidates and matching to 4F operators\label{sect:models}}

In this section we will count the number of UV completions for 4F
operators at 1-loop level and describe their matter
content. Furthermore, we will investigate and discuss some patterns among the SMEFT operators that these models generate and present the explicit matching for particular examples.

\begin{table}[ht!]
    \centering
    \begin{multicols}{2}
    \hspace*{-0.2cm}
    \begin{tabular}{|l|l|l|}
        \hline
        \textbf{Class} & \textbf{Name} & \textbf{Structure} \\
        \hline \hline
        LL & $\mathcal{O}_{ll}$ &  $ \left(\bar{l}_L \gamma_{\mu} l_L\right) \left(\bar{l}_L \gamma^{\mu} l_L\right)$ \\
        \hline
        & $\mathcal{O}_{le}$ & $ \left(\bar{l}_L \gamma_{\mu} l_L \right)\left(\bar{e}_R \gamma^{\mu} e_R \right)$ \\
        \hline
        & $\mathcal{O}_{ee}$ & $ \left(\bar{e}_R \gamma_{\mu} e_R \right) \left(\bar{e}_R \gamma^{\mu} e_R \right)$ \\
        \hline \hline
        LQ & $\mathcal{O}_{lq}^{(1)}$ & $\left(\bar{l}_L \gamma_{\mu} l_L\right) \left(\bar{q}_L \gamma^{\mu} q_L\right)$ \\
        &$\mathcal{O}_{lq}^{(3)}$ &  $\left(\bar{l}_L \gamma_{\mu} \sigma_a l_L\right) \left(\bar{q}_L \gamma^{\mu} \sigma_a q_L\right)$ \\
        \hline
        & $O_{lu}$ & $ \left(\bar{l}_L \gamma_{\mu} l_L\right) \left(\bar{u}_R \gamma^{\mu} u_R\right)$ \\
        \hline
        &  $\mathcal{O}_{ld}$ & $\left(\bar{l}_L \gamma_{\mu} l_L\right)\left(\bar{d}_R \gamma^{\mu} d_R\right) $ \\
        \hline
        & $\mathcal{O}_{lequ}^{(1)}$ &  $\left(\bar{l}_L e_R \right) i \sigma_2 \left(\bar{q}_L u_R \right)^T$ \\
        & $\mathcal{O}_{lequ}^{(3)}$ & $\left(\bar{l}_L \sigma_{\mu\nu} e_R \right) i \sigma_2 \left(\bar{q}_L \sigma^{\mu \nu} u_R \right)^T$ \\
        \hline
        & $\mathcal{O}_{ledq}$ & $\left(\bar{l}_L e_R \right) \left(\bar{d}_R q_L \right)$ \\
        \hline
        & $\mathcal{O}_{qe}$ & $\left(\bar{q}_L \gamma_{\mu} q_L\right) \left(\bar{e}_R \gamma^{\mu} e_R\right)$ \\
        \hline
        & $\mathcal{O}_{eu}$ & $\left(\bar{e}_R \gamma_{\mu} e_R\right) \left(\bar{u}_R \gamma^{\mu} u_R\right)$ \\
        \hline
        & $\mathcal{O}_{ed}$ & $\left(\bar{e}_R \gamma_{\mu} e_R\right) \left(\bar{d}_R \gamma^{\mu} d_R\right)$ \\
        \hline
   \end{tabular}
   
   \columnbreak
   
   \begin{tabular}{|l|l|l|}
        \hline
        \textbf{Class} & \textbf{Name} & \textbf{Structure} \\
        \hline \hline
        QQ & $\mathcal{O}_{qq}^{(1)}$ & $\left(\bar{q}_L \gamma_{\mu} q_L\right) \left(\bar{q}_L \gamma^{\mu} q_L\right)$ \\
        &  $\mathcal{O}_{qq}^{(3)}$ & $\left(\bar{q}_L \gamma_{\mu} \sigma_a q_L\right) \left(\bar{q}_L \gamma^{\mu} \sigma_a q_L\right)$ \\
        \hline 
        & $\mathcal{O}_{quqd}^{(1)}$ & $\left(\bar{q}_L u_R \right) \left(\bar{q}_L d_R \right)^T$ \\
        & $\mathcal{O}_{quqd}^{(8)}$ & $\left(\bar{q}_L T_A u_R \right) \left(\bar{q}_L T_A d_R \right)^T$ \\
        \hline
        & $\mathcal{O}_{qu}^{(1)}$ & $\left(\bar{q}_L \gamma_{\mu} q_L\right) \left(\bar{u}_R \gamma^{\mu} u_R\right)$ \\
        & $\mathcal{O}_{qu}^{(8)}$ & $\left(\bar{q}_L \gamma_{\mu} T_A q_L\right) \left(\bar{u}_R \gamma^{\mu} T_A u_R\right)$ \\
        \hline 
        & $\mathcal{O}_{qd}^{(1)}$ & $\left(\bar{q}_L \gamma_{\mu} q_L\right)\left(\bar{d}_R \gamma^{\mu} d_R\right)$ \\
        & $\mathcal{O}_{qd}^{(8)}$ & $\left(\bar{q}_L \gamma_{\mu} T_A q_L\right) \left(\bar{d}_R \gamma^{\mu} T_A d_R\right)$ \\
        \hline
        & $\mathcal{O}_{uu}$ & $ \left(\bar{u}_R \gamma_{\mu} u_R\right) \left(\bar{u}_R \gamma^{\mu} u_R\right)$ \\
        \hline
        & $\mathcal{O}_{ud}^{(1)}$ & $\left(\bar{u}_R \gamma_{\mu} u_R\right) \left(\bar{d}_R \gamma^{\mu} d_R\right)$ \\
        & $\mathcal{O}_{ud}^{(8)}$ & $\left(\bar{u}_R \gamma_{\mu} T_A u_R\right) \left(\bar{d}_R \gamma^{\mu} T_A d_R\right)$ \\
        \hline
        & $\mathcal{O}_{dd}$ & $\left(\bar{d}_R \gamma_{\mu} d_R\right) \left(\bar{d}_R \gamma^{\mu} d_R\right)$ \\
        \hline
    \end{tabular}
    \end{multicols}
    \caption{List of baryon (and lepton) number conserving 4F
      operators in the Warsaw basis at $d =$6. Note that we have
      suppressed generation indices here.}
    \label{tab:4-fermion_operators}
\end{table}

\begin{table}[]
    \centering
    \begin{tabular}{|l|l|l|}
        \hline
        \textbf{Class} & \textbf{Name} & \textbf{Structure} \\
        \hline \hline
        $\psi^2 \phi^2 D$ & $\mathcal{O}_{H l}^{(1)}$ & $\left( H^{\dagger} i \overset{\leftrightarrow}{D}_{\mu} H\right) \left(\bar{l}_L \gamma^{\mu} l_L\right)$ \\
        & $\mathcal{O}_{H l}^{(3)}$ & $\left( H^{\dagger} i \overset{\leftrightarrow}{D}_{\mu}^a H\right) \left(\bar{l}_L \gamma^{\mu} \sigma^a l_L\right)$ \\
        \hline
        & $\mathcal{O}_{H e}$ & $\left( H^{\dagger} i \overset{\leftrightarrow}{D}_{\mu} H\right) \left(\bar{e}_R \gamma^{\mu} e_R \right)$ \\
        \hline
        & $\mathcal{O}_{H q}^{(1)}$ &  
        $\left( H^{\dagger} i \overset{\leftrightarrow}{D}_{\mu} H\right) \left(\bar{q}_L \gamma^{\mu} q_L\right)$ \\
        & $\mathcal{O}_{H q}^{(3)}$ & $\left( H^{\dagger} i \overset{\leftrightarrow}{D}_{\mu}^a H\right) \left(\bar{q}_L \gamma^{\mu} \sigma^a q_L\right)$ \\
        \hline
        & $\mathcal{O}_{H u}$ & $\left( H^{\dagger} i \overset{\leftrightarrow}{D}_{\mu} H\right) \left(\bar{u}_R \gamma^{\mu} u_R \right)$ \\
        \hline
        & $\mathcal{O}_{H d}$ & $\left( H^{\dagger} i \overset{\leftrightarrow}{D}_{\mu} H\right) \left(\bar{d}_R \gamma^{\mu} d_R \right)$ \\
        \hline
        & $\mathcal{O}_{H u d}$ & $\left( H^{\dagger} i \overset{\leftrightarrow}{D}_{\mu} H\right) \left(\bar{u}_R \gamma^{\mu} d_R \right)$ \\
        \hline
        \hline
        $\psi^2 \phi^3$ &  $\mathcal{O}_{eH}$ & $\left( H^{\dagger} H\right)^2 \left(\bar{l}_L H e_R \right)$ \\
        \hline
        & $\mathcal{O}_{uH}$ & $\left( H^{\dagger} H\right)^2 \left(\bar{q}_L H u_R \right)$ \\
        \hline
        & $\mathcal{O}_{dH}$ & $\left( H^{\dagger} H\right)^2 \left(\bar{q}_L H d_R \right)$ \\
        \hline
    \end{tabular}
    \caption{Fermion-Higgs operators at dimension-6 in the Warsaw
      basis.}
    \label{tab:2fnh_operators}
\end{table}

\subsection{Counting UV models}
\label{subsect:counting_models}

In the following we will focus on flavour-diagonal 4F operators with
no baryon number violation. While the model
diagrams allow for open flavour indices, we restrict ourselves to the
flavour-diagonal case because experimental constraints for
flavour-violating operators are much stronger than their
flavour-conserving counter part. This leaves us, in the Warsaw basis, 
with 25 4F operators at dimension 6, as listed in Table
\ref{tab:4-fermion_operators}.

Furthermore, for simplicity, we will just distinguish
operators by their external fields, for example we simply write $\mathcal{O}_{lq}$ for
both $\mathcal{O}_{lq}^{(1)}$ and $\mathcal{O}_{lq}^{(3)}$.  
Hence, we are left with 18 B-conserving 4F operators which can be
classified into 3 categories:

\begin{enumerate}
\item 3 lepton-specific operators (dubbed LL): $\mathcal{O}_{ll}$, $\mathcal{O}_{le}$ and $\mathcal{O}_{ee}$
\item 7 quark-specific operators (QQ): $\mathcal{O}_{qq}$, $\mathcal{O}_{quqd}$, $\mathcal{O}_{qu}$, $\mathcal{O}_{qd}$, $\mathcal{O}_{uu}$, $\mathcal{O}_{ud}$, $\mathcal{O}_{dd}$
\item 8 mixed lepton-quark operators (LQ): $\mathcal{O}_{lq}$, $\mathcal{O}_{lu}$, $\mathcal{O}_{ld}$, $\mathcal{O}_{lequ}$, $\mathcal{O}_{leqd}$, $\mathcal{O}_{qe}$, $\mathcal{O}_{eu}$, $\mathcal{O}_{ed}$
\end{enumerate}

We use the \textit{ModelGenerator}, a Mathematica code that implements the described diagrammatic approach, to find all UV models that open up any of these 18 operator structures at one-loop and meet the criteria discussed in the previous section. In particular this means that we search only for box diagrams with $Z_2$-odd BSM scalars and fermions, as motivated in section \ref{subsect:diagrammatic_approach}, and containing at least one of the DM candidates introduced in section \ref{subsect:DM_content}.

The procedure to list possible models following this diagrammatic approach is general and would lead to an infinite number of models if no restrictions on the possible particle content were imposed. But since in both discussed cases the list of either exits or DM candidates is finite, we are left with a finite number of models. As this number can still be large it will prove useful to limit the number of models and define sub-classes of models according to their particle content.

For models containing exit particles, these sub-classes could be identified by choosing certain limits for the representations under $SU(3)_C$, $SU(2)_L$ and $U(1)_Y$, e.g. by focusing only on $SU(2)$ doublets instead of going all the way up to $SU(2)$ quadruplets. Here, 1-loop models containing DM candidates can be straight-forwardly classified according to the \textit{type} of DM candidate, distinguishing between those already described in section~\ref{subsect:DM_content}.

\begin{itemize}
\item DM candidates with hypercharge $Y=0$, i.e. $(S/F)_{1,n,0}$ with $n = 1, 3, 5, ...$. We will look only at models for the singlet and triplet cases here. Nevertheless, the number of models for higher multiplets coincides with the number of models for the triplet case by construction and the new particle content can be reconstructed easily from the triplet case, as it will be explained in subsection \ref{subsect:new_particles}. 

\item For $Y=\frac{1}{2}$ only the scalar DM candidates could exist on their own even in the absence of NRO terms for mass splitting. We will consider the inert doublet DM model featuring $S_{1,2,1/2}$ as an example here.
 
\item Finally, DM candidates with $Y=1$ will not be taken into account here.
\end{itemize}

Table \ref{tab:DM_candidates} summarises the choices and nomenclature for DM candidates we will be considering.

\begin{table}[t]
  \begin{center}
      \begin{tabular}{|l|l|}
      	\hline
		\textbf{Name} & \textbf{DM candidates} \\  
      	\hline
      	\hline
      	$Y=0$ singlets & $S_{1,1,0}$, $F_{1,1,0}$ \\
      	\hline
      	$Y=0$ triplets & $S_{1,3,0}$, $F_{1,3,0}$ \\
      	\hline
      	inert doublet & $S_{1,2,1/2}$ \\
      	\hline 
      \end{tabular}
    \caption{DM candidates considered in the counting of UV models for 4F operators at 1-loop.}
    \label{tab:DM_candidates}
  \end{center}
\end{table}

\subsubsection{Overlap between models for different 4F operators}

Before studying concrete models it is worth to have a look at the model overlap between different 4F operators. Depending on the structure of the operator, a box model contains between two and four different BSM particles, one of them being a DM candidate. If a subset of these particles provides a model for any other 4F operator we say that this model creates an overlap between the two operators. To quantify this co-generation of different operators we count the model overlap for every 4F operator with every other of the 17 4F operator structures in a so-called overlap matrix.

\begin{figure}[t!]
    \centering
    \includegraphics[scale=0.5]{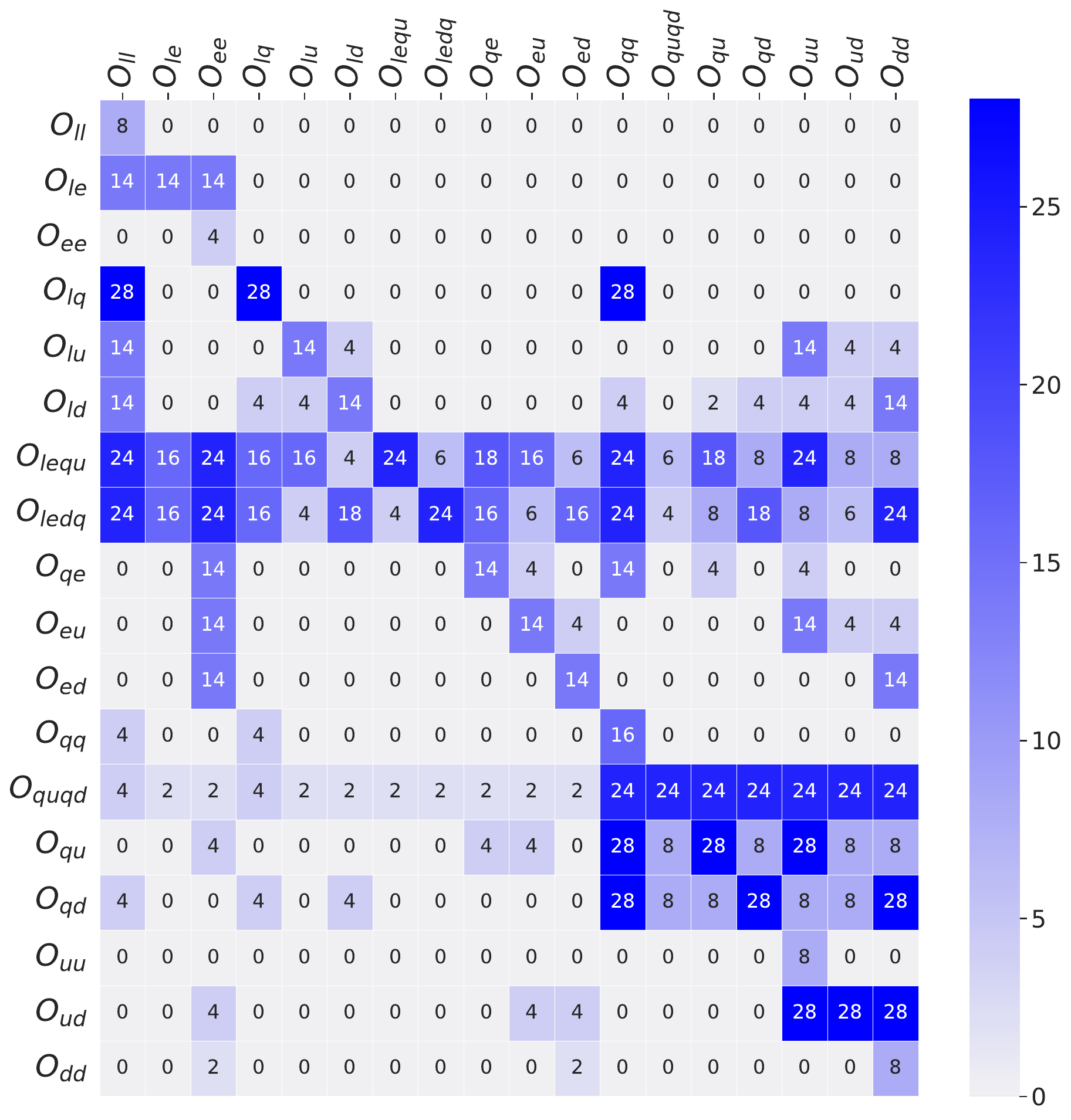}
    \caption{Operator overlap matrix for models with a singlet dark
      matter candidates. The entries on the diagonal count the number
      of models for the given operator. The entry in row $i$ and column $j$ 
      corresponds to the number of models for operator ${\cal O}_i$ that also
      generate operator ${\cal O}_j$.}
    \label{fig:overlap_matrix_singlet}
\end{figure}

\begin{figure}[t!]
    \centering
    \includegraphics[scale=0.5]{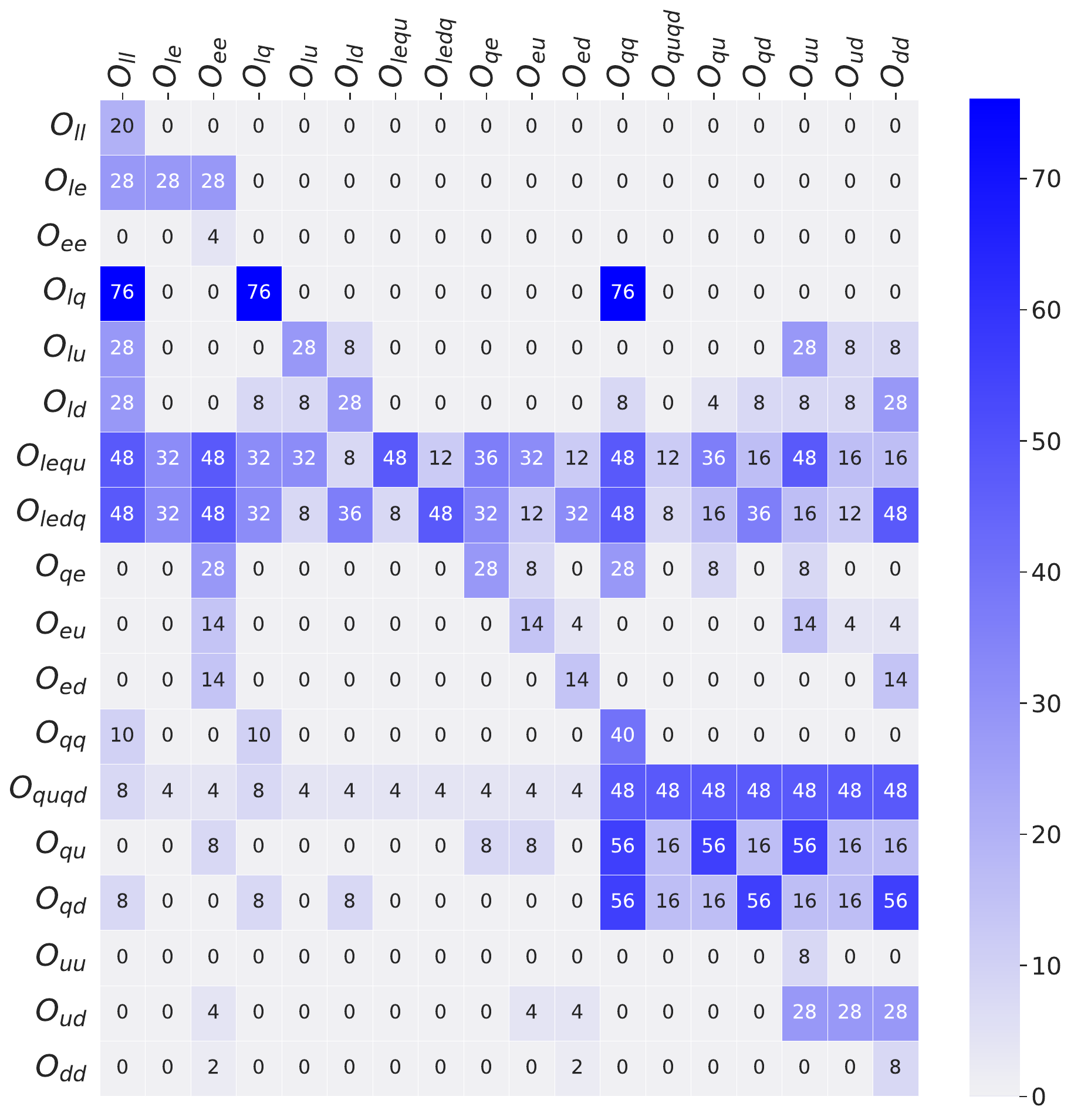}
    \caption{As Fig.\ \ref{fig:overlap_matrix_singlet} but for
      triplets and higher multiplets as DM candidates.}
    \label{fig:overlap_matrix_triplet}
\end{figure}

Firstly, starting with $Y=0$ DM candidates, Fig.\ \ref{fig:overlap_matrix_singlet} shows the overlap matrix for singlets and Fig.\ \ref{fig:overlap_matrix_triplet} for triplets and higher representations, respectively. To emphasise the large hierarchy between the entries in the matrix, we underlie the table with a heat-map, with darker colours representing larger numbers.

The table is to be read in the following way: The diagonal entries list the total number of models for each operator. Every row belongs to one operator $\mathcal{O}_X$. Then, every entry in this row lists how many of the models for operator $\mathcal{O}_X$ contain a sub-model that contributes to the operator $\mathcal{O}_Y$, which labels the column. Note that we do not require that the box diagram which generates operator $\mathcal{O}_Y$ in the overlap matrix to contain a DM candidate. However, by construction, all particles in the loop must be odd under the stabilising symmetry.

\begin{figure}
    \begin{subfigure}[b]{0.5\textwidth}
       \centering
       \includegraphics[width=\textwidth]{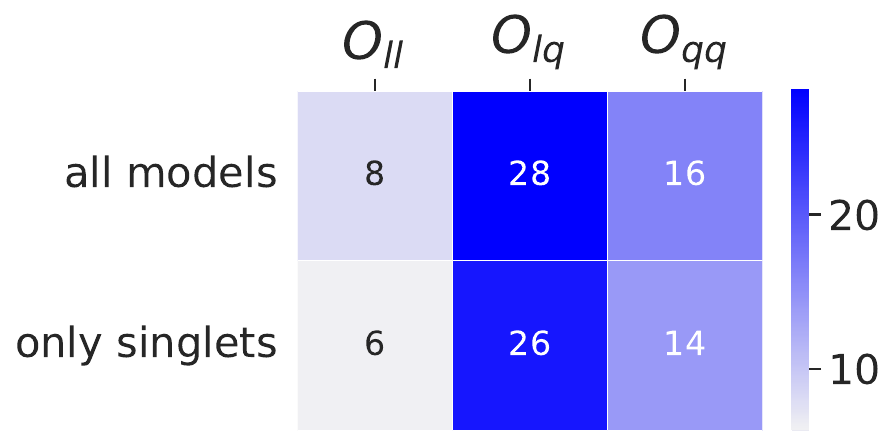}
	\subcaption{For singlets}
    \end{subfigure}
    \begin{subfigure}[b]{0.5\textwidth}
	    \centering
	    \includegraphics[width=\textwidth]{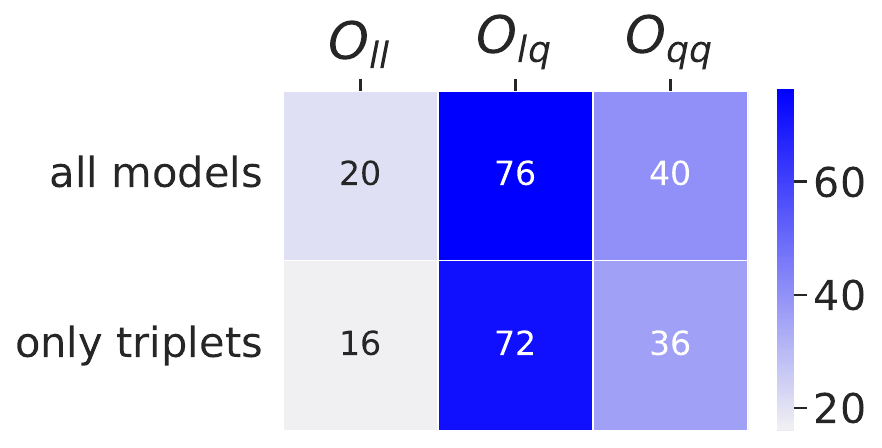}
	\subcaption{For triplets}
    \end{subfigure}
    \caption{Number of models for the three operators $\mathcal{O}_{ll}$, $\mathcal{O}_{lq}$ and $\mathcal{O}_{qq}$, in the first row considering all models that contain DM $Y=0$ singlets (left) or triplets (right). The second row counts the models that do not contain any other DM multiplet. For details see text.}
    \label{fig:overlap_one_multiplet_only}
\end{figure}

As a last comment, it should be mentioned that the above overlap matrices only require the particular DM multiplet, singlet or triplet, to be present. Thus, the list of models will also contain examples of models with more than one viable DM candidate. If we want to ensure that the models contain only one DM $Y=0$ candidate, we can exclude all other multiplets by hand. This does not change much the number of models, though, and the reason is fairly simple: To have e.g. $SU(2)$ singlets and triplets present at the same time in one loop requires four external $SU(2)$ doublets. This is only the case for the three operators $\mathcal{O}_{ll}$, $\mathcal{O}_{lq}$ and $\mathcal{O}_{qq}$. For the singlet case, the diagonal entries for these operators will be reduced by two, as we exclude the two models that contain $S_{110}$ and $S_{130}$, or $F_{110}$ and $F_{130}$, see Fig.\ \ref{fig:overlap_one_multiplet_only} (left). For triplets, the diagonal entries are reduced by four, since we now have to exclude the models with $S_{110}$ and $S_{130}$, $F_{110}$ and $F_{130}$, $S_{130}$ and $S_{150}$, $F_{130}$ and $F_{150}$, see Fig.\ \ref{fig:overlap_one_multiplet_only} (right). Again, higher multiplets lead to the exactly same numbers as for the triplet case.

\begin{figure}[t!]
    \centering
    \includegraphics[scale=0.5]{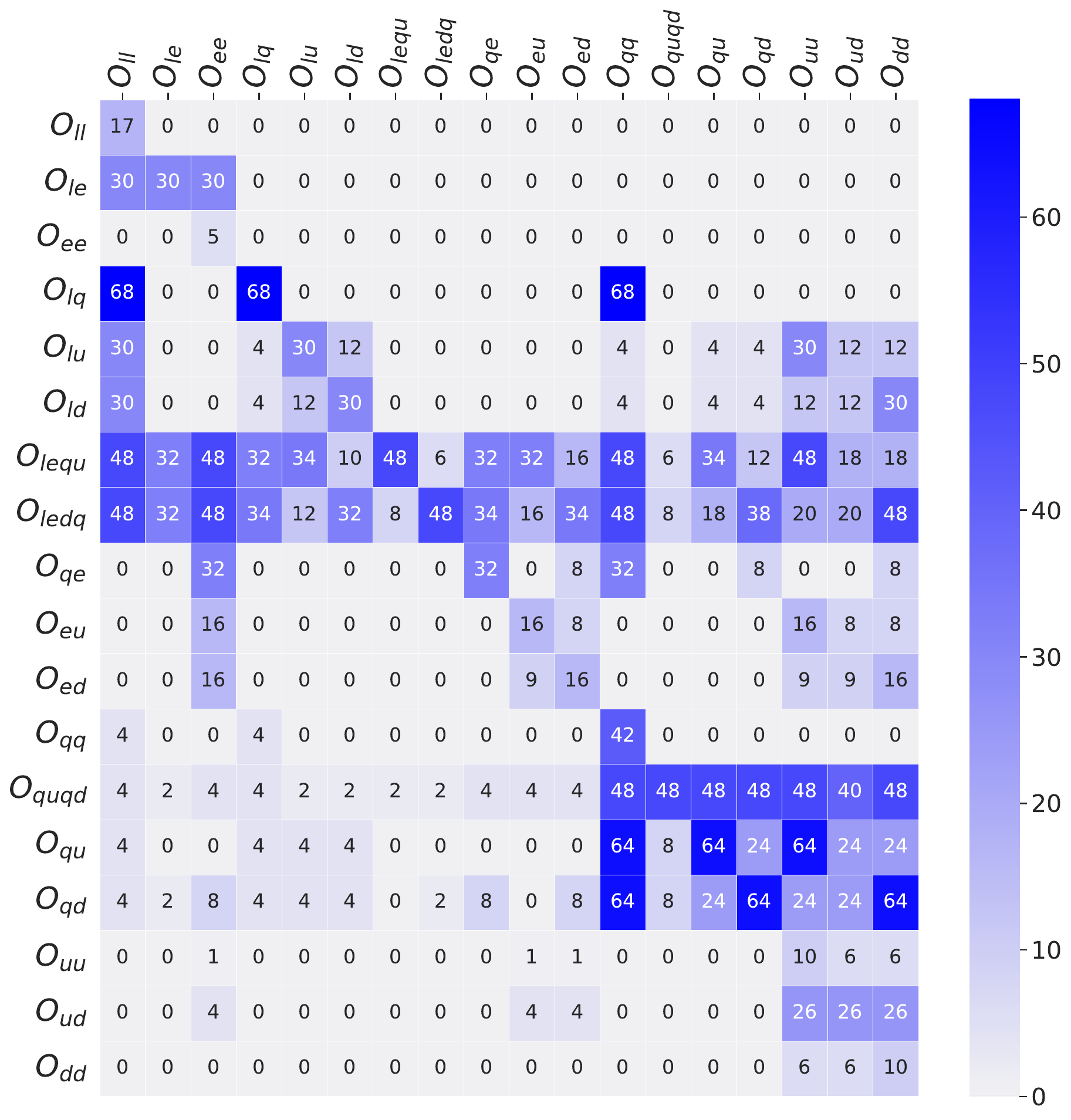}
    \caption{Operator overlap for $S_{1,2,1/2}$.}
    \label{fig:overlap_matrix_S1212}
\end{figure}

Finally, Fig.\ \ref{fig:overlap_matrix_S1212} shows the model overlap featuring $S_{1,2,1/2}$ as a DM candidate. Note that considering instead a DM candidate $S_{1,n,1/2}$ with $n=4,6, ...$ would not change much the numbers. Only the number of models for operators with four external $SU(2)$ doublets is slightly larger for $n\neq 2$, since these models can contain particles with an $SU(2)$ representation of $n-2$ that did not appear for $S_{1,2,1/2}$.

\subsubsection{Overlap with Fermion-Higgs operators}

Some of the models that generate 4F operators at 1-loop will unavoidably open up operators with external fermions and Higgs bosons of the type $\psi^2 \phi^2 D$ and $\psi^2 \phi^3$, see Table \ref{tab:2fnh_operators}. In the following we will refer to both of these classes of operators together as fermion-Higgs operators (FH).

\begin{figure}[t!]
    \centering
    \includegraphics[scale=0.55]{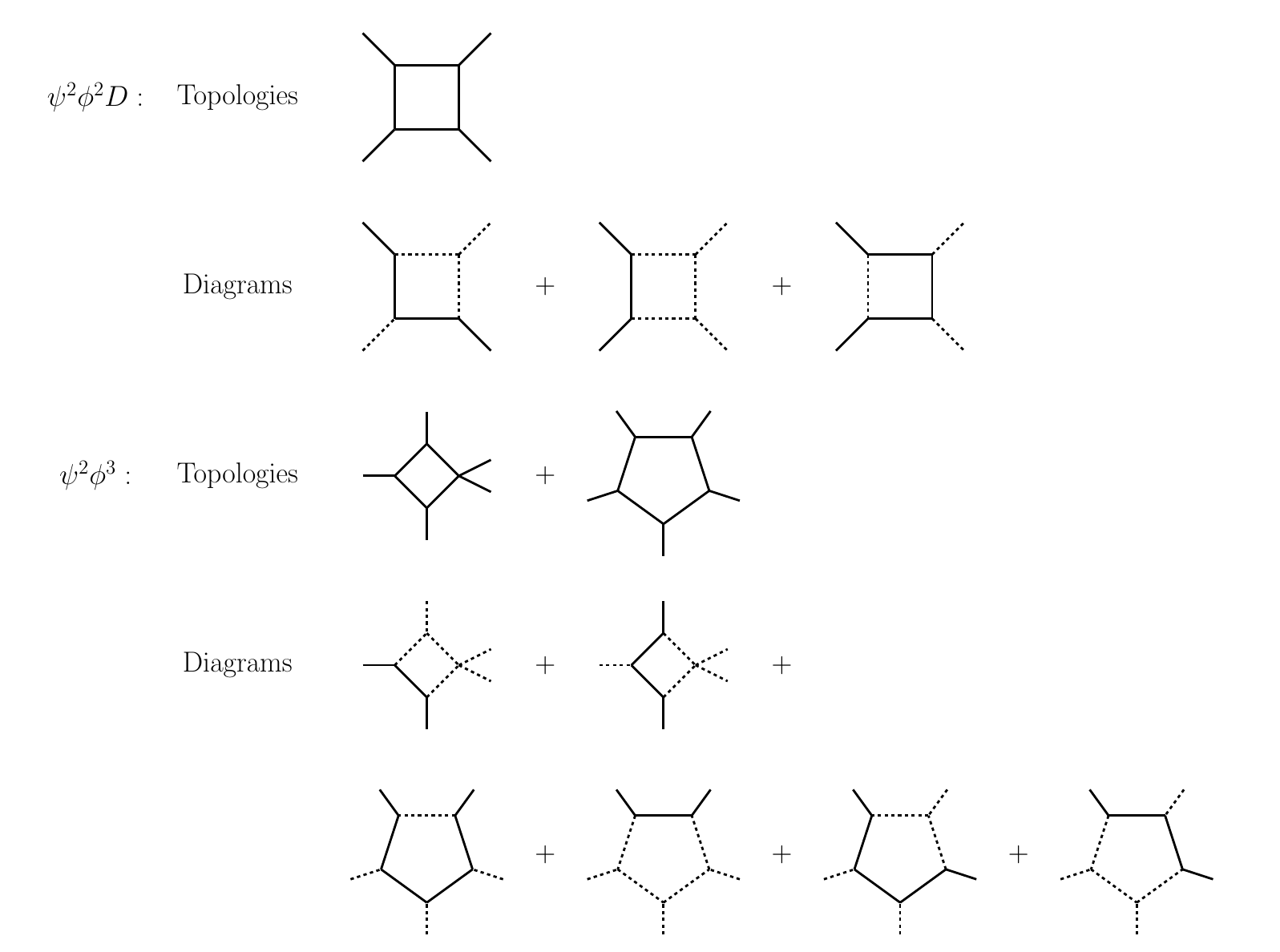}
    \caption{Topologies and diagrams for the operators $\psi^2 \phi^2 D$ and $\psi^2 \phi^3$.}
    \label{fig:topologies_FH}
\end{figure}

In contrast to the case with exit particles in the loop, the contributions to FH operators from models with DM candidates will only appear at 1-loop, but not at tree-level. This follows from the fact that the new particles running in the loop for 4F operators are odd under the stabilising $Z_2$ symmetry and can thus not couple at tree-level to a pair of $Z_2$-even SM particles. Fig.\ \ref{fig:topologies_FH} shows the possible topologies and diagrams for the two classes of FH operators. Note that here we have given only the topologies that lead to \textit{proper} model-diagrams for FH operators:  The diagrams for the operators containing one derivative, $\psi^2 \phi^2 D$, can at 1-loop level only be generated by box diagrams, similar to the boxes for 4-fermion operators. On the other hand, the model diagrams for the operators with 2 external fermions and three Higgses, $\psi^2 \phi^3$, can be generated either via pentagram diagrams or boxes with a scalar four-point vertex.  As discussed in Sec.\ \ref{subsect:diagrammatic_approach}, we consider only diagrams with internal scalars and fermions, but no vectors. 
  
\begin{figure}[t!]
    \centering
    \includegraphics[scale=0.5]{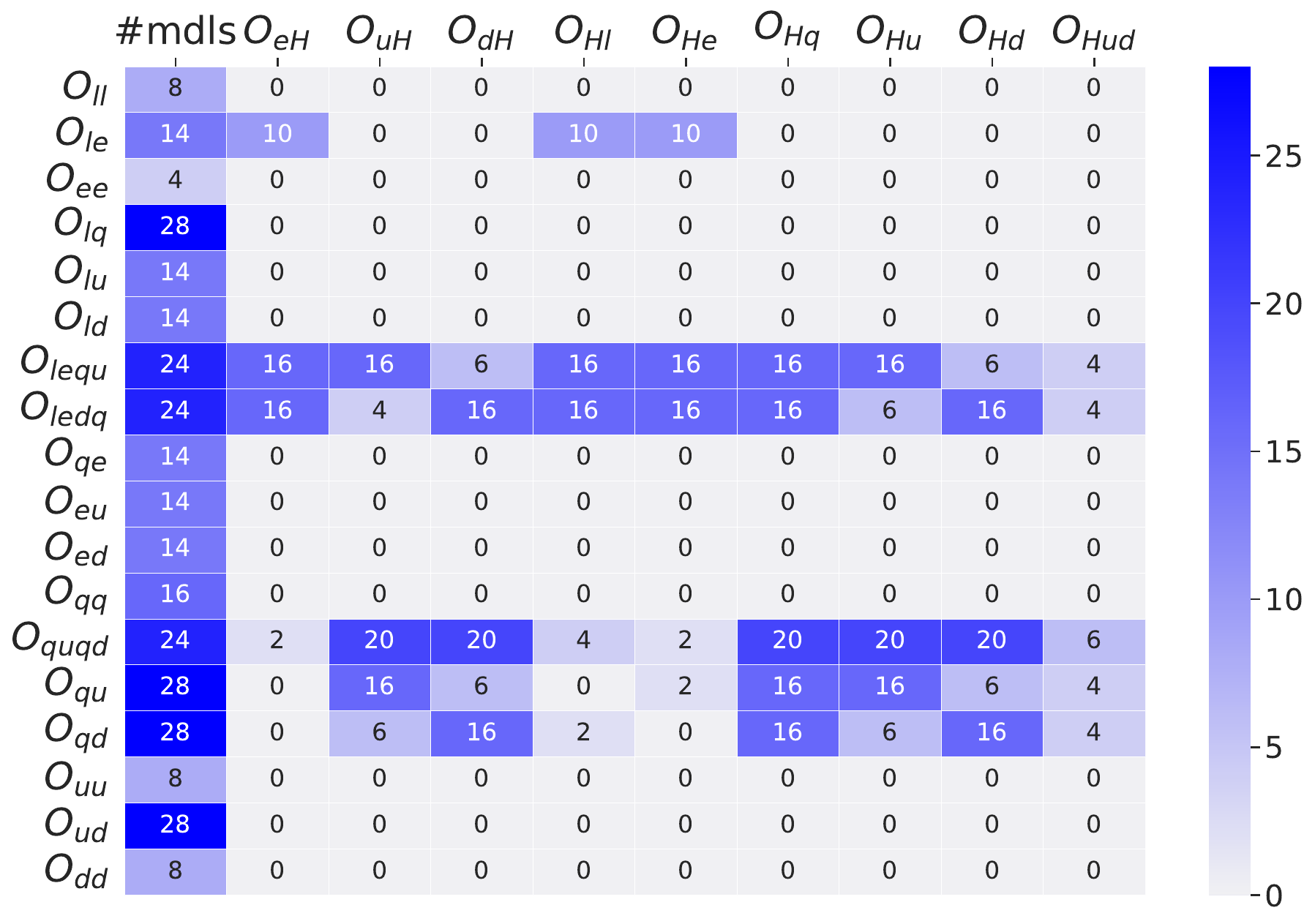}
    \caption{Operator overlap with fermion-Higgs operators for singlets.}
    \label{fig:overlap_matrix_FH_singlet}
\end{figure}

\begin{figure}[t!]
    \centering
    \includegraphics[scale=0.5]{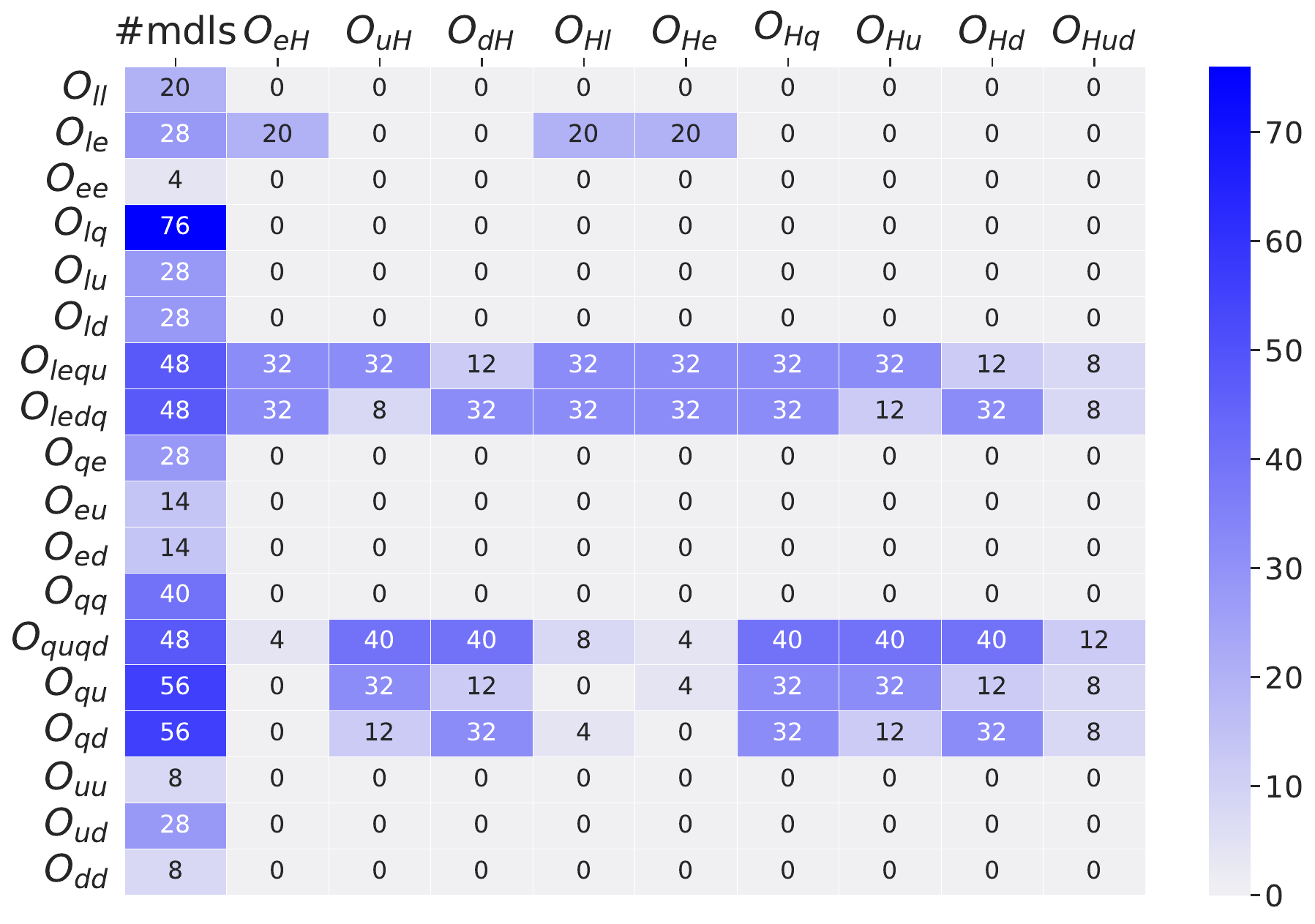}
    \caption{Operator overlap with fermion-Higgs operators for triplets.}
    \label{fig:overlap_matrix_FH_triplet}
\end{figure}

\begin{figure}[t!]
    \centering
    \includegraphics[scale=0.5]{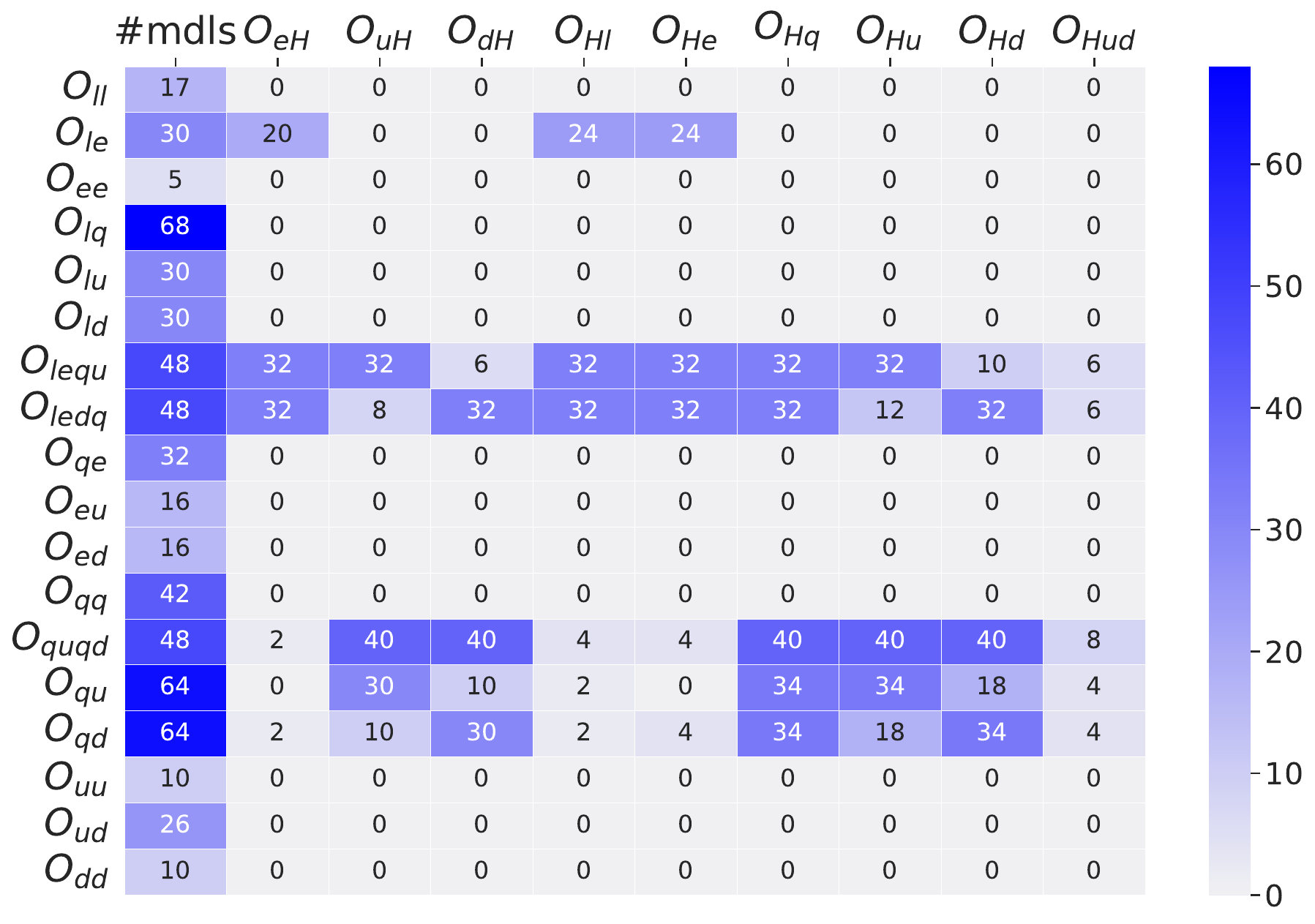}
    \caption{Operator overlap with fermion-Higgs operators for $S_{1,2,1/2}$.}
    \label{fig:overlap_matrix_FH_S1212}
\end{figure}

Fig.\ \ref{fig:overlap_matrix_FH_singlet} shows the overlap between 4F and FH operators for DM $Y=0$ singlets, Fig.\ \ref{fig:overlap_matrix_FH_triplet} for DM $Y=0$ triplets and Fig.\ \ref{fig:overlap_matrix_FH_S1212} for the inert doublet model featuring $S_{1,2,1/2}$. Which operator will actually be opened by each model depends on the
nature of the new fermions though: Models with Dirac BSM fermions will
open up different operators than models with Majorana BSM fermions,
see the discussion in section \ref{subsect:match}. The overlap
matrices do not specify the nature a priori and contain entries for all
operators that could be opened for either Dirac or Majorana
fermions.
 
Matching the 1-loop models to the relevant SMEFT operators, the Wilson coefficients for the FH operators will in general be of the same order as the Wilson coefficients for the 4F operators since now both 4F and FH operators are produced at 1-loop. However, as we will discuss in section \ref{subsect:match}, for particular points in the parameter space of masses and couplings of the new particles, the FH operators can provide the dominant contribution. Hence it is important to keep in mind which models for four-fermion operators can open up models for fermion-Higgs operators and thus may be constrained by the latter.

\subsection{Patterns in 4F operators}
\label{subsect:patterns}

The overlap matrices Fig.\ \ref{fig:overlap_matrix_singlet}, \ref{fig:overlap_matrix_triplet} and \ref{fig:overlap_matrix_S1212} feature only particular DM candidates, but still contain a large amount of models per operator. Nevertheless, having a closer look at the tables a few general patterns can be recognised. As long as we consider both the scalar and fermionic DM candidate for a given $Y=0$ multiplet, all models are completely symmetric under the exchange of BSM fermions and scalars. As one can see in table \ref{tab:DM_candidates}, every scalar DM candidate has a fermionic partner with the same quantum numbers and vice versa. Starting the diagrammatic approach from either a scalar or fermionic DM candidate as seed will lead to models with the same quantum numbers for each particle, swapping fermions and scalars. Accordingly, all the numbers appearing in the model and overlap counts are necessarily even and it would actually suffice to generate one half of them (e.g.\ starting only from scalar DM seeds) and complete the other half by exchanging scalars for fermions. This symmetry can be spotted in the tables \ref{tab:new_particles_singlet} and \ref{tab:new_particles_triplet} for the particle content too.

However, this symmetry will of course be broken if we consider only a particular scalar or fermionic DM
candidate, as we did for the inert doublet model with
$S_{1,2,1/2}$. The following asymmetry in scalars and fermions in the
models can be observed in Fig.\ \ref{fig:overlap_matrix_S1212}, as now
odd entries appear and directly in Table \ref{tab:new_particles_S1212}.

Note that this symmetry could not be found in the first place when considering models with exits because exit partners with different spin-statistics are not always an exit themselves.

\begin{figure}
    \begin{subfigure}[b]{0.48\textwidth}
        \centering
        \includegraphics[width=0.8\textwidth]{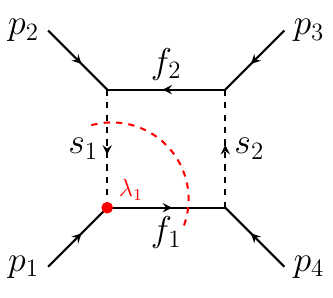}
        \subcaption{Generic 4F operator with up to four different external legs.}
	\end{subfigure}
	\hfill
    \begin{subfigure}[b]{0.48\textwidth}
        \centering
        \includegraphics[width=0.8\textwidth]{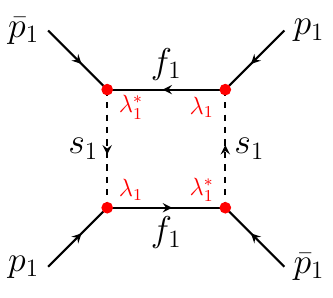}
        \subcaption{Generic 4F operator with all four legs being the same SM particle.}
	\end{subfigure}
    \caption{Schematic example how to obtain models for \textit{more symmetric} 4F operators from models for \textit{less symmetric} ones. A model that generates an operator of the form $\mathcal{O}_{p_1 p_2 p_3 p_4}$ will automatically produce also the operators $\mathcal{O}_{p_1 p_1}$, $\mathcal{O}_{p_2 p_2}$, $\mathcal{O}_{p_3 p_3}$ and $\mathcal{O}_{p_4 p_4}$. See text for details.}
    \label{fig:OLQvsOLL}
\end{figure}

Also it is worth noticing that the overlap matrices Fig.\ \ref{fig:overlap_matrix_singlet}, \ref{fig:overlap_matrix_triplet} and \ref{fig:overlap_matrix_S1212} are not symmetric under the exchange of rows and columns, i.e.\ the number of models for operator $\mathcal{O}_X$ that open up $\mathcal{O}_Y$ and the number of models for operator $\mathcal{O}_Y$ that open up $\mathcal{O}_X$ are different. This can be simply seen from the different structure of the operators. An example is illustrated in Fig.\ \ref{fig:OLQvsOLL}: Taking a model for an operator with different external legs (in this case $p_1, p_2, p_3, p_4$), cutting out one corner and stitching it together in a box with copies of itself (conjugated when necessary), will lead to a model for every operator $\mathcal{O}_{p_i}$ for any of $p_1$, $p_2$, $p_3$ and $p_4$. So all models for $\mathcal{O}_{p_1 p_2 p_3 p_4}$ overlap with all $\mathcal{O}_{p_i}$, but not vice versa.

To give a concrete example, focus on the leptonic sector for $Y=0$ singlets, i.e.\ the first three rows and columns in Fig.\ \ref{fig:overlap_matrix_singlet}. All of the 14 models for $\mathcal{O}_{le}$ will generate models for $\mathcal{O}_{ll}$ and $\mathcal{O}_{ee}$ by the procedure just described. On the other hand, none of the 8 models for $\mathcal{O}_{ll}$, nor the 4 models for $\mathcal{O}_{ee}$, open up any model for $\mathcal{O}_{le}$. Note that this last result also differs from the findings for models with exits: In the latter case, some models for $\mathcal{O}_{ll}$ could open $\mathcal{O}_{le}$ by allowing SM particles in the box, in this case the Higgs doublet. This is not possible for any DM model since we require a $Z_2$ symmetry that stabilises the DM candidate, thus all particles in the box must be odd under this symmetry and no SM particle can contribute in the boxes.

Third, it is striking that the matrices are in general sparse and contain many zeros. This means that many models contribute only to a subset of operators. Hence, a good model discrimination can be achieved if some Wilson coefficients are experimentally established to be non-zero. 
Given this sparseness and to connect to phenomenology, it is convenient to group the models in three sub-classes:
\begin{itemize}
    \item {\bf Lepton-specific scenarios:} exclusively producing lepton-specific (LL) 4F operators.
    \item {\bf Quark-specific scenarios:} exclusively producing quark-specific (QQ) 4F operators.
    \item {\bf Generic or hybrid scenarios:} which contribute to the three types (LL, QQ and hybrid) of 4F operators.
\end{itemize} 
While lepton-specific scenarios are well-constrained from low-energy precision measurements, quark-specific scenarios will provide a better setting for direct searches at the LHC.

With these definitions and the overlap matrices at hand, we can readily classify the models that we found. To start with, no model for lepton-specific operators ($\mathcal{O}_{ll}, \mathcal{O}_{le}$ and $\mathcal{O}_{ee}$) generates any model for quark-specific or mixed quark-lepton operators. This holds for all DM candidates considered here ($Y=0$ singlets and triplets and $S_{1,2,1/2}$), as can be seen from the zero entries in the first three lines and the columns 4-18 in Figs.\ \ref{fig:overlap_matrix_singlet}, \ref{fig:overlap_matrix_triplet} and \ref{fig:overlap_matrix_S1212}. Again, in contrast to exit models, the reason for this is the absence of SM particles in the boxes due to the $Z_2$ symmetry. Accordingly, all models for lepton-specific operators lead directly to lepton-specific scenarios.

The same statement does not hold for quark-specific models. Since the quark operators involve up to 3 different external particles ($q$, $u$ and $d$), the total number of models is larger and a few will contribute to lepton-specific and hybrid operators. However, the number of these models is still small in comparison with the total number of models. Therefore, many quark-specific models can be identified by selecting the models for QQ operators that do not produce any LQ or LL overlap. Nevertheless, there is one caveat to have in mind: If an operator that produces only 4F operators that fall in the class QQ opens up any lepton-Higgs operator, i.e.\ $\mathcal{O}_{H l}^{(1)}$, $\mathcal{O}_{H l}^{(3)}$, $\mathcal{O}_{H e}$ or $\mathcal{O}_{e H}$ in Table \ref{tab:2fnh_operators}, the low-energy constraints from the latter will dominate over the actual constraints from QQ 4F operators. So to fully classify a model as quark-specific requires to also exclude any contribution to lepton-Higgs operators, which are provided in Fig.\ \ref{fig:overlap_matrix_FH_singlet}, \ref{fig:overlap_matrix_FH_triplet} and \ref{fig:overlap_matrix_FH_S1212}.

In summary, the overlap matrices for 4F operators in DM models are 
much sparser than what is found for ``exit'' models. This implies 
that there is a \textit{high} model discrimination power in 4F operators, if any 
of these could be measured in some future experiment.

\subsection{New particles in 1-loop models for 4F operators}
\label{subsect:new_particles}

Having counted possible UV models and classified them according to the
operators they contribute to, it is time to look at what new particles
are actually contained in these type of models.

Table \ref{tab:new_particles_singlet} lists all the particles that
appear in models for the B-conserving 4F operators that contain a DM
$Y=0$ singlet $S_{110}$ or $F_{110}$. Table
\ref{tab:new_particles_triplet} does the same for DM $Y=0$ triplets
$S_{130}$ or $F_{130}$. Note that the particle content for models with
higher DM multiplets can be easily obtained by replacing the $SU(2)$
representations of all the particles. For example, the complete set of
loop particles for models with DM $Y=0$ quintuplets is obtained from
Table \ref{tab:new_particles_triplet} replacing singlets by triplets,
doublets by quadruplets, triplets by quintuplets, quadruplets by
sextets and quintuplets by septuplets under $SU(2)$. Table
\ref{tab:new_particles_S1212} lists the particles that can appear in
boxes for models that contain the inert doublet $S_{1,2,1/2}$. Note
that some representations appear only as scalars because their
fermionic partner would require to include $F_{1,2,1/2}$ as a valid DM
candidate, leading to a scalar-fermion asymmetry in the possible
particle content.

As a closing comment to this section, we note that for all 
BSM models introducing new scalars and/or fermions, the running 
of SM parameters at high energies will be changed, potentially 
leading to the appearance of new Landau poles below the Planck 
scale. Excluding the existence of such Landau poles could be considered 
a model building criterion. However, since we are mostly interested 
in models that give phenomenology at LHC energies, we did not 
study RGE running of SM parameters for our model lists.

\begin{table}[h]
\centering
\begin{tabular}{|l|l|l|l|l|}
\hline
\backslashbox{$SU(2)$}{$SU(3)$} & singlets & triplets & sextets & octets \\
\hline \hline
singlets & $ 0, 1, 2$ & $-\frac{1}{3}, \frac{2}{3}, -\frac{4}{3}, \frac{5}{3}$ & $\frac{1}{3}, -\frac{2}{3}, \frac{4}{3}$ & $0, 1$ \\ 
\hline
doublets & $\frac{1}{2}, \frac{3}{2}$ & $\frac{1}{6}, -\frac{5}{6}, \frac{7}{6}$ & $-\frac{1}{6}, \frac{5}{6}$ & $\frac{1}{2}$ \\
\hline 
triplets & $0, 1$ & $-\frac{1}{3}, \frac{2}{3}$ & $\frac{1}{3}$ & $0$ \\
\hline
\end{tabular}
\caption{The quantum numbers for both new scalars and fermions that appear in the boxes for models for 4F operators which contain a DM $Y=0$ singlet $S_{110}$ or $F_{110}$. Each cell lists the possible values of the hypercharge $Y$ that are found for the given combination of $SU(2)$ and $SU(3)$ representations.}
\label{tab:new_particles_singlet}
\end{table}

\begin{table}[h]
\centering
\begin{tabular}{|l|l|l|l|l|}
\hline
\backslashbox{$SU(2)$}{$SU(3)$} & singlets & triplets & sextets & octets \\
\hline \hline
singlets & $ 0, 1$ & $-\frac{1}{3}, \frac{2}{3}$ & $\frac{1}{3}$ & $0$ \\ 
\hline
doublets & $\frac{1}{2}, \frac{3}{2}$ & $\frac{1}{6}, -\frac{5}{6}, \frac{7}{6}$ & $-\frac{1}{6}, \frac{5}{6}$ & $\frac{1}{2}$ \\
\hline 
triplets & $0, 1, 2$ & $-\frac{1}{3}, \frac{2}{3}, -\frac{4}{3}, \frac{5}{3}$ & $\frac{1}{3}, -\frac{2}{3}, \frac{4}{3}$ & $0, 1$ \\
\hline
quadruplets & $\frac{1}{2}, \frac{3}{2}$ & $\frac{1}{6}, -\frac{5}{6}, \frac{7}{6}$ & $-\frac{1}{6}, \frac{5}{6}$ & $\frac{1}{2}$ \\
\hline
quintuplets & $0, 1$ & $-\frac{1}{3}, \frac{2}{3}$ & $\frac{1}{3}$ & $0$ \\
\hline
\end{tabular}
\caption{The quantum numbers (possible values of $Y$ for given $SU(2)$ and $SU(3)$ representation) for both new scalars and fermions that appear in the boxes for models for 4F operators which contain a DM $Y=0$ triplet $S_{130}$ or $F_{130}$. }
\label{tab:new_particles_triplet}
\end{table}

\begin{table}[h]
\centering
\begin{tabular}{|l|l|l|l|l|}
\hline
\backslashbox{$SU(2)$}{$SU(3)$} & singlets & triplets & sextets & octets \\
\hline \hline
singlets & $ \mathbf{0}, \mathbf{1}, \mathit{2}$ & $\mathbf{-\frac{1}{3}}, \mathbf{\frac{2}{3}}, \mathit{-\frac{4}{3}}, \mathit{\frac{5}{3}}$ & $\mathit{\frac{1}{3}}, \mathit{-\frac{2}{3}}, \mathit{\frac{4}{3}}$ & $\mathit{0}, \mathit{1}$ \\ 
\hline
doublets & $\mathbf{\frac{1}{2}}, \mathbf{\frac{3}{2}}, \mathit{\frac{5}{2}}$ & $\mathbf{\frac{1}{6}}, \mathbf{-\frac{5}{6}}, \mathbf{\frac{7}{6}}, \mathit{-\frac{11}{6}}, \mathit{\frac{13}{6}}$ & $\mathit{-\frac{1}{6}}, \mathit{\frac{5}{6}}, \mathit{-\frac{7}{6}}, \mathit{\frac{11}{6}}$ & $\mathit{\frac{1}{2}},\mathit{\frac{3}{2}} $ \\
\hline 
triplets & $ \mathbf{0}, \mathbf{1}, \mathit{2}$ & $\mathbf{-\frac{1}{3}}, \mathbf{\frac{2}{3}}, \mathit{-\frac{4}{3}}, \mathit{\frac{5}{3}}$ & $\mathit{\frac{1}{3}}, \mathit{-\frac{2}{3}}, \mathit{\frac{4}{3}}$ & $\mathit{0}, \mathit{1}$ \\ 
\hline
quadruplets & $\mathit{\frac{1}{2}}, \mathit{\frac{3}{2}}$ & $\mathit{\frac{1}{6}}, \mathit{-\frac{5}{6}}, \mathit{\frac{7}{6}}$ & $\mathit{-\frac{1}{6}}, \mathit{\frac{5}{6}}$ & $\mathit{\frac{1}{2}}$ \\
\hline \hline
\end{tabular}
\caption{The quantum numbers (possible values of $Y$ for given $SU(2)$ and $SU(3)$ representation) for both new scalars and fermions that appear in the boxes for models for 4F operators which contain the DM candidate $S_{1,2,1/2}$. The representations in bold are found for both scalars and fermions, the italic representations only for scalars.}
\label{tab:new_particles_S1212}
\end{table}

\subsection{Matching of specific models\label{subsect:match}}

In this section we will discuss the matching of some specific DM
models. For the 1-loop matching we use \verb|MatchmakerEFT|
\cite{Carmona:2021xtq}. For all model variants discussed below, 
we have written \verb|FeynRules| \cite{Alloul:2013bka} files, 
implementing the models in the unbroken phases (since we are 
matching to the SMEFT). These model files can be found in the supplementary material.\\
We separate the following discussion into two groups: (i) Models 
that generate leptonic operators and (ii) Models that 
generate operators with quarks.

\subsubsection{Leptophilic models\label{subsubsect:Scot}}

Perhaps the simplest DM model that can generate ${\cal O}_{ll}$ is 
a SM extension with two fields: $F_{1,1,0}$ and $S_{1,2,1/2}$. 
This model (with three copies of $F_{1,1,0}$) is known in 
the literature as the ``scotogenic'' model \cite{Ma:2006km}. This 
DM model has the additional motivation that it can explain not 
only DM, but also neutrino masses and their mixing angles.
The most important terms in the Lagrangian of this model are:
\begin{equation}\label{eq:lagsc}
{\cal L}^{\rm Sc} \propto Y_{\nu} {\bar N_R}L \eta + \frac{1}{2} \lambda_5 (H\eta^{\dagger})^2 
+ \text{h.c.} + \cdots \, ,
\end{equation}
where $N_R\equiv F_{1,1,0}$ and $\eta\equiv S_{1,2,1/2}$, in the
original notation, and considering  $N_R$ as a Majorana field. The model generates the Weinberg operator at 1-loop level.\footnote{The model assumes a $Z_2$ under which both $N_R$ and $\eta$ 
are odd, as well as that $\eta$ remains ``inert'' after electro-weak 
symmetry breaking, i.e. $\langle \eta^0 \rangle=0$.} In the limit of $\Lambda=m_N=m_\eta$ 
its coefficient (in one-generation notation) is simply given by 
$c_W = -\frac{\lambda_5}{16 \pi^2}\frac{Y_{\nu}^2}{4 \Lambda}$. 

We will call this model Sc-I in the following, short for ``scotogenic
type-I''. One can easily find variants of this model, for example one
can replace $N_R\equiv F_{1,1,0}$ by $\Sigma\equiv F_{1,3,0}$, which
we will call Sc-III. One could also extend Sc-I by adding a second
scalar, $S_{1,1,1}$; we will call this variant Sc-I$^+$. In the
scotogenic model the fermion(s) is(are) assumed to be Majorana
particles. For reasons to be discussed below, we also introduce Sc-I
and Sc-III assuming the fermions to be Dirac particles, Sc-I-D and
Sc-III-D, respectively.  We have implemented all these models in
\verb|MatchmakerEFT| \cite{Carmona:2021xtq} with one generation of
heavy fermions.\footnote{Performing a neutrino oscillation fit would
  require at least two copies of fermions. Such a fit could be done
  easily, see for example \cite{Cordero-Carrion:2019qtu}.  Since in
  this paper we are interested in $d=6$ operators, we do not repeat
  this discussion.}

\begin{figure}[t]
\begin{center}
\includegraphics[width=0.9\linewidth]{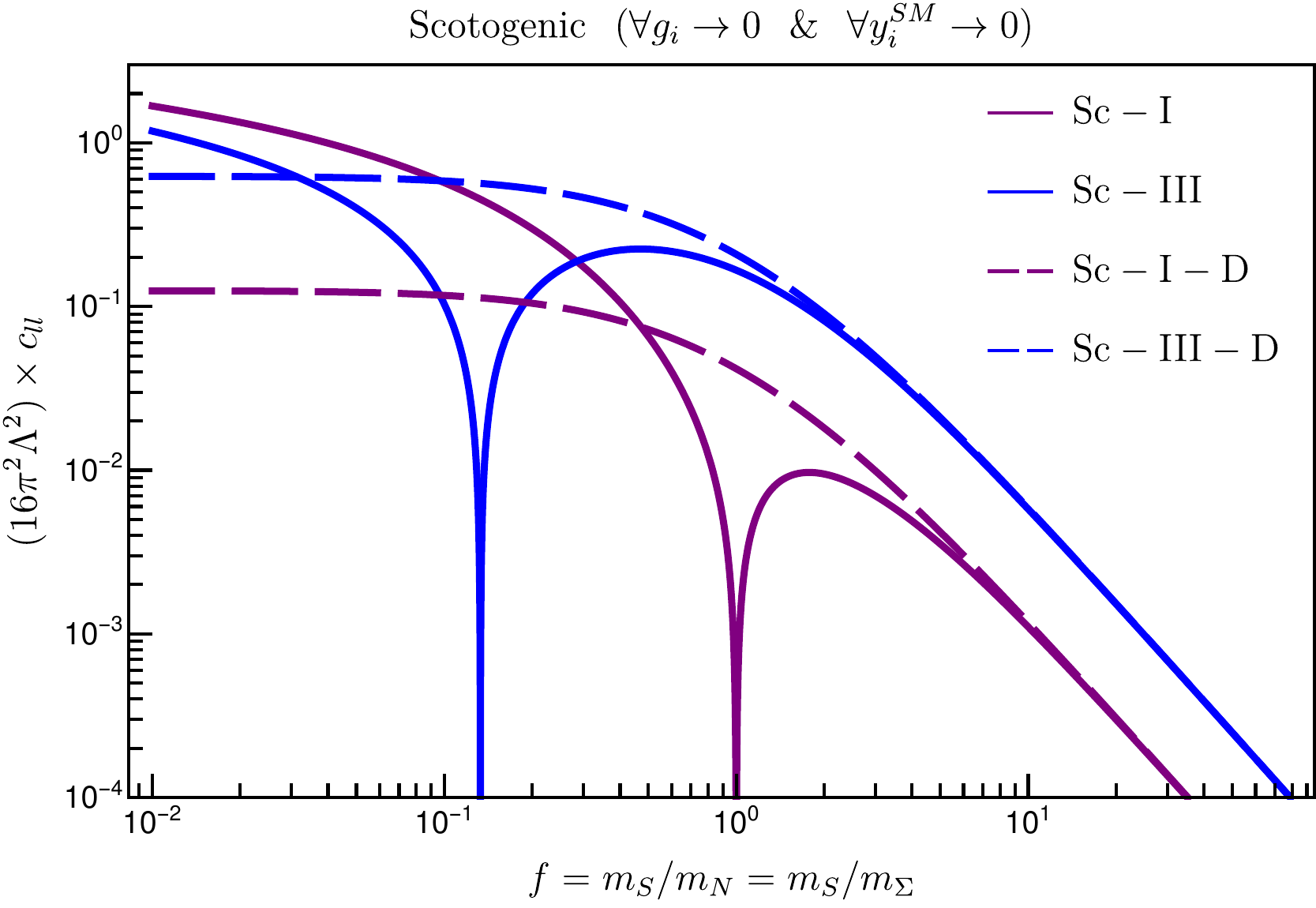}
\caption{Matching of the coefficient of ${\cal O}_{ll}$ for different
  variants of the ``scotogenic'' model, assuming the Yukawa couplings
  equal to one and neglecting SM couplings. We multiply by a 
factor $(16 \pi^2\Lambda^2)$, with $\Lambda=m_N$ or $m_\Sigma$ 
for the type-I and type-III models, respectively. }
\label{fig:OllMtch1}
\end{center}
\end{figure}

Fig.\ \ref{fig:OllMtch1} shows the results for the coefficient
$c_{ll}$ as a function of $f=m_S/m_N=m_S/m_{\Sigma}$ for four
different model variants in the limit of SM couplings going to zero,
i.e. $g_{1,2}\to 0$.  We plot $c_{ll}(16 \pi^2\Lambda^2)$ as a
function of $f$, so the function shown is independent 
of the overall scaling of the $d=6$ operator, ($1/\Lambda^2$).

An interesting feature occurs in the two Majorana models. For Sc-I, at
$f=1$ the coefficient $c_{ll}$ vanishes.  We have traced this back to
an exact cancellation among different diagrams, that is due to the
Majorana nature of the fermion (note that for Sc-III the same
cancellation occurs, but at a value of $f$ roughly $f\simeq 0.13$). 
We note that the same Majorana cancellation has been found 
in a different context in \cite{Arnan:2019uhr}. For a Majorana 
fermion, there are always two types of diagrams, one with a 
lepton number conserving and one with a lepton number violating 
fermion propagator. This two diagrams come with opposite signs, 
and lead to the observed cancellation, for some specific value 
of the scalar mass(es) in the diagram.
To demonstrate that this cancellation is indeed caused by the Majorana
nature of the fermion, we also show the result for the models Sc-I-D
and Sc-III-D. For the Dirac models the coefficient $c_{ll}$ never
vanishes, as Fig.\ \ref{fig:OllMtch1} demonstrates.

In the limit where all SM couplings are neglected and for $f=1$, one
can write the coefficient $c_{ll}$ as
\begin{equation}\label{eq:cll}
(c_{ll})_{\alpha,\beta,\gamma,\delta}^{X} = \frac{\tilde{c}_{ll}^X}{16 \pi^2\Lambda^2}
Y_{\alpha}Y_{\gamma}Y^{*}_{\beta}Y^{*}_{\delta},
\end{equation}
where $Y$ stands symbolically for the BSM Yukawa coupling (either
type-I or type-III) and the coefficient $\tilde{c}_{ll}^X$ is given in
Table \ref{tab:coefficients_leptophilic} for the different model
variants.

\begin{table}[h]
\centering
\begin{tabular}{|c|c|c|c|c|}
\hline
model X & Sc-I & Sc-I-D & Sc-III & Sc-III-D \\
\hline
\hline
$\tilde{c}_{ll}^{X}$ & 0 & $-\frac{1}{24}$ & $-\frac{1}{6}$ & $-\frac{5}{24}$ \\
\hline
\end{tabular}
\caption{$\tilde{c}_{ll}$ for different models for $f \rightarrow 1$
  and neglecting all SM couplings. For details see text.}
\label{tab:coefficients_leptophilic}
\end{table}

\begin{figure}[t]
\begin{center}
\includegraphics[width=0.9\linewidth]{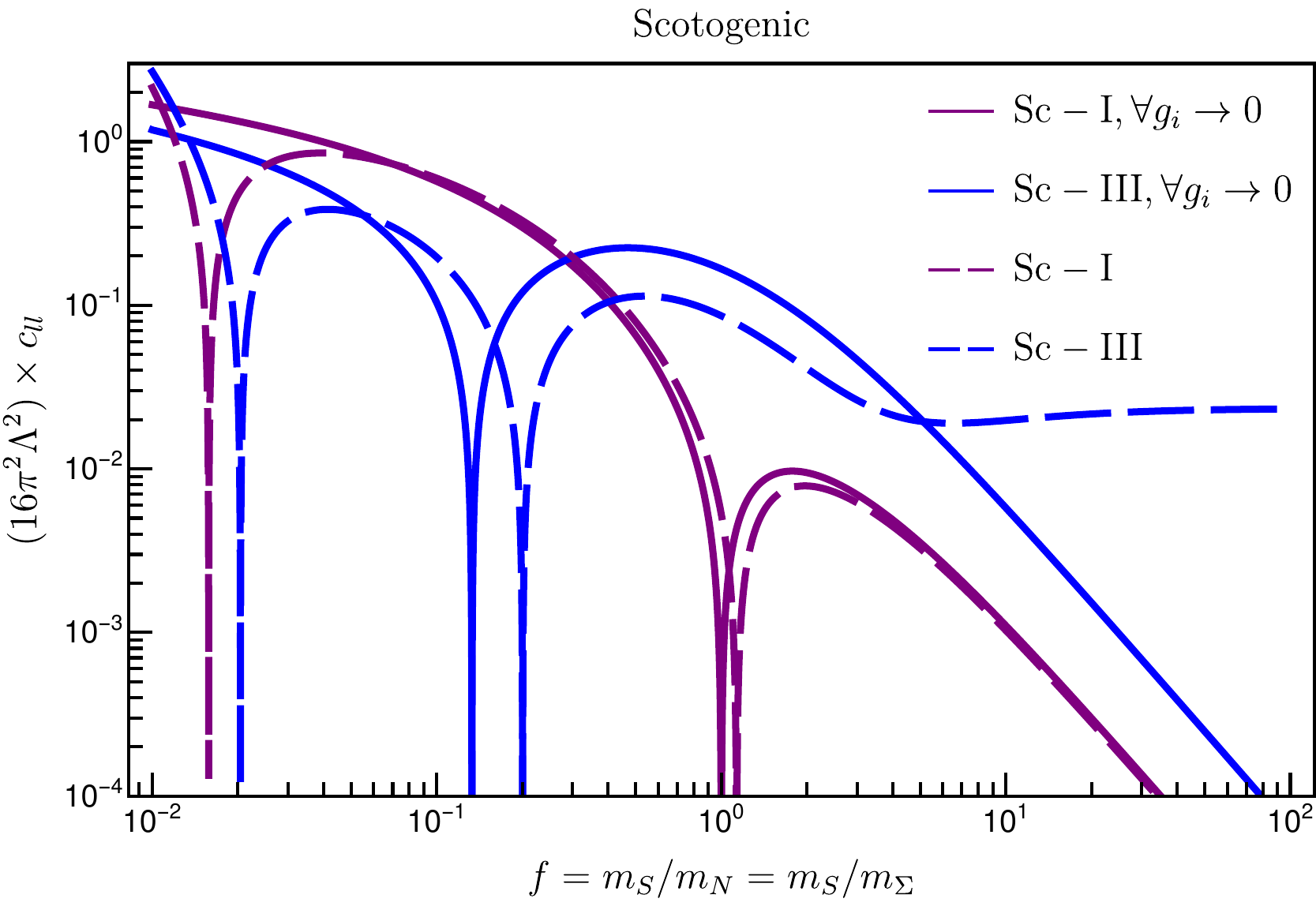}
\cprotect\caption{Matching of the coefficient of ${\cal O}_{ll}$ for
  different variants of the ``scotogenic'' model. Full lines neglect
  SM couplings, dashed line is the full result from
  \verb|MatchmakerEFT|. In all cases we assume Yukawa couplings to be
  equal to one. $\Lambda=m_N = m_\Sigma$. }
\label{fig:OllMtch2}
\end{center}
\end{figure}

Fig.\ \ref{fig:OllMtch2} compares the matching results for the 
Majorana variants for vanishing SM couplings and for the full 
results, as found by \verb|MatchmakerEFT|. For the Sc-I model, 
the contributions from the Yukawa couplings (proportional to 
$Y^4$) dominates nearly everywhere, except for a region at very 
small values of $f$, where a second cancellation among diagrams 
appear at very small values of $f$. Note also that taking into 
account non-zero gauge couplings shift the cancellation slightly 
away from $f=1$. Non-zero gauge couplings are more important in the 
case of Sc-III. This is easily understood: $F_{1,1,0}$ in type-I 
has no gauge couplings, while $F_{1,3,0}$ in type-III contributes 
to the matching via $(D_{\mu}W^{\mu\nu})^2$ with a sizeable  
coefficient. Also for Sc-III a second cancellation region appears 
at small $f$. Noteworthy is also that for $g_i\to 0$, the coefficient $c_{ll}$ goes 
to zero in the limit of large $f$. This is not the case for 
Sc-III when $g_2$ is switched on, again due to the $F_{1,3,0}$ coupling to 
the electro-weak current. 

In the limit of vanishing SM couplings, the four model variants discussed so far contribute only to 
${\cal O}_{ll}$. Contributions 
to other 4F operators are always of the form $(g_i^{\rm SM})^2(y^{SM}_l)^2$, $(g_i^{\rm SM})^2Y^2$ or $(g_i^{\rm SM})^4$. Numerically these contributions 
are usually much smaller than those of $c_{ll}\propto Y^4$ \footnote{Here and in the following we use $Y$ symbolically to denote any BSM 
Yukawa coupling.} and we 
will not discuss them in detail here.

\begin{figure}[t]
\begin{center}
\includegraphics[width=0.9\linewidth]{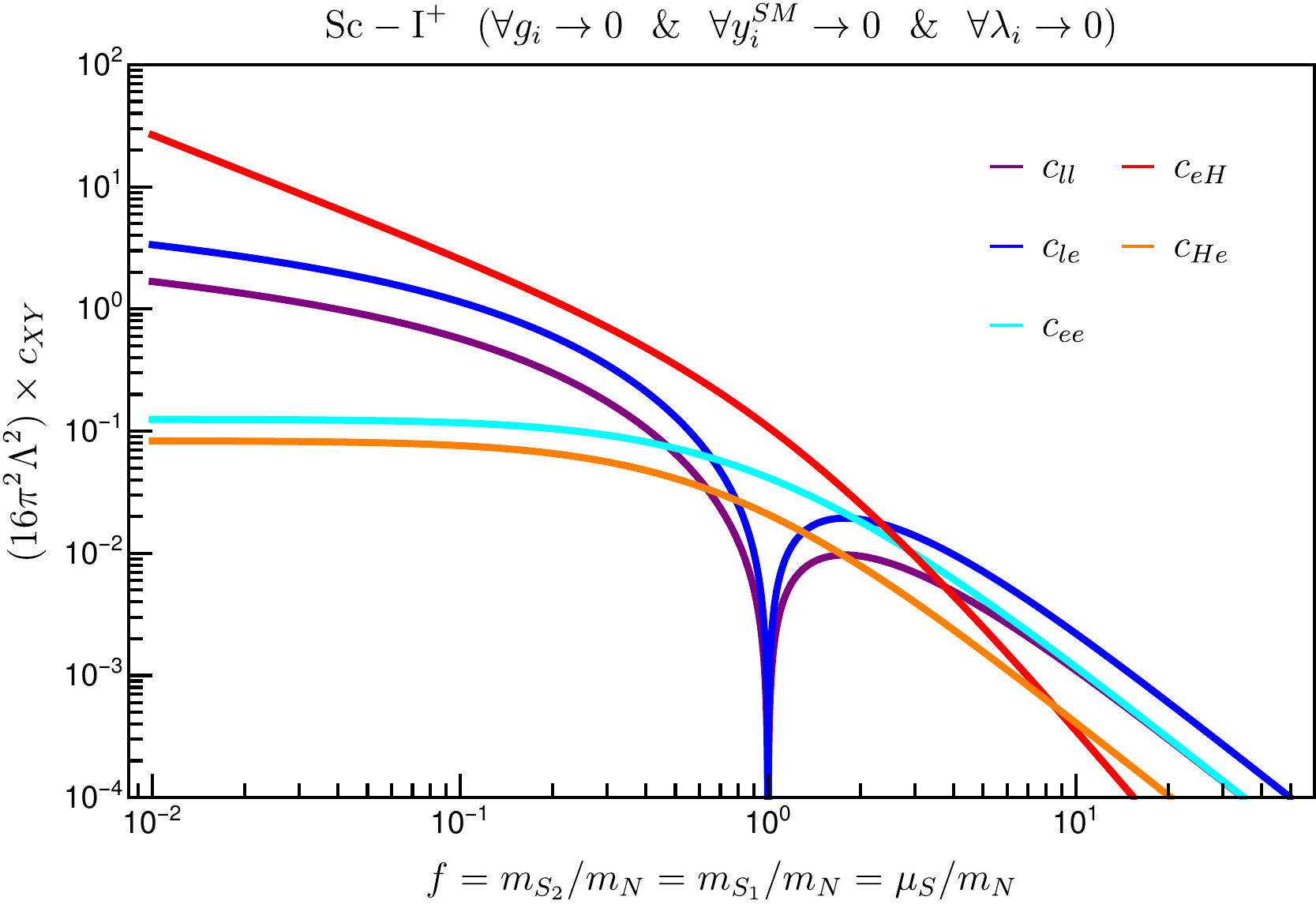}
\caption{Matching of different coefficients $c_{XY}$ in the 
model Sc-I$^+$. The results have been calculated in the limit 
of vanishing SM couplings and for BSM Yukawas equal to one. $\Lambda=m_N$.}
\label{fig:OllMtch3}
\end{center}
\end{figure}

The model variant Sc-I$^+$, featuring a second scalar $S_{1,1,1}$, adds a second Yukawa coupling $Y_{NE} \, {\bar
  F^c_{1,1,0}} \, e_R \, S_{1,1,1}$ to the model Sc-I. We have constructed
this variant because it gives a much richer set of non-zero
coefficients for 4F and FH operators, as
Fig.\ \ref{fig:OllMtch3} shows. The coefficient $c_{ll}$ in this model coincides with
$c_{ll}$ of Sc-I, but in addition the model produces non-zero
coefficients for $c_{le}$ and $c_{ee}$. Note that both $c_{ll}$ and
$c_{le}$ show the Majorana cancellation at $f=1$, but $c_{ee}$ does
not. Neither do the coefficients for the operators including Higgses.

\begin{table}[h]
\centering
\begin{tabular}{|l|c|c|c|c|c|c|c|}
\hline
XY & $ll$ & $le$ & $ee$ & $Hl^{(1)}$ & $Hl^{(3)}$ & $He$ & $eH$ \\
\hline
\hline
$\tilde{c}_{XY}$ & 0 & 0 & - $\frac{1}{24}$ & $-\frac{1}{96}$ & $-\frac{1}{96}$ & $\frac{1}{48}$ & $\frac{1}{4}$ \\
\hline
\end{tabular}
\caption{Limits for $f \rightarrow 1$ of the matching results for different Wilson coefficients ${\tilde{c}_{XY} = (16 \pi^2 \Lambda^2) \times c_{XY}}$  in the model Sc-I$^+$. We consider BSM Yukawas equal to one.}
\label{tab:limits_f1_ScI+}
\end{table}

For completeness we mention the values
for the matching for $f=1$ for the different operators in Table \ref{tab:limits_f1_ScI+}. 
These values are valid in the limit where possible BSM
quartic couplings are put to zero. We note that we have also checked
that if all $Y=1$, contributions from SM gauge and
Yukawa couplings lead only to minor corrections to the matching
coefficients show in Fig.\ \ref{fig:OllMtch3}, for $f\gsim 0.1$.

\subsubsection{Coloured models\label{subsubsect:Col}}

We now turn to the matching for some example models including coloured
fields. While there are many possible variants, one can examine the
main aspects of these types of models, as far as $d=6$ EFT operators
are concerned, just by discussing the following three models.
One of the simplest models that can be constructed with
coloured fields takes the scotogenic model and replaces the scalar
$S_{1,2,1/2}$ by $S_{3,2,1/6}$. This simple variant, with $F_{1,1,0}$,
called CM1 in the following, will generate a box diagram for the
operator ${\cal O}_{qq}^{(1)}$ (instead of ${\cal O}_{ll}$ as in the
scotogenic model). Similarly, as discussed for the leptophilic models 
above, we can replace  $F_{1,1,0}$ by  $F_{1,3,0}$ to arrive at a 
model we will call CM3. Finally, one can add $F_{1,2,1/2}$ to CM1 
to construct a model which we will call CM2.

We note in passing that the models CM1 and CM3 produce diagrams which are
essentially the same as the one-loop diagrams one encounters in the
MSSM. This is, of course, by construction, since $S_{3,2,1/6}$ is
equivalent to the scalar quark doublet of SUSY, ${\tilde Q}_L$, while
$F_{1,1,0}$ and $F_{1,3,0}$ correspond to the bino and wino of
supersymmetry. However, in the following we will consider the
couplings connecting these BSM fields to the SM (and to each other) as
free ``Yukawa'' couplings. In a supersymmetric world these couplings
would be fixed by gauge couplings instead. Also, note that different
from the leptophilic models, the coloured models discussed here do
not have any connection with neutrino masses.\footnote{Extensions 
which have both lepton and quark $d=6$ operators (and generate neutrino 
masses) can, of course, be straightforwardly constructed.}

\begin{figure}[t]
\begin{center}
\includegraphics[width=\linewidth]{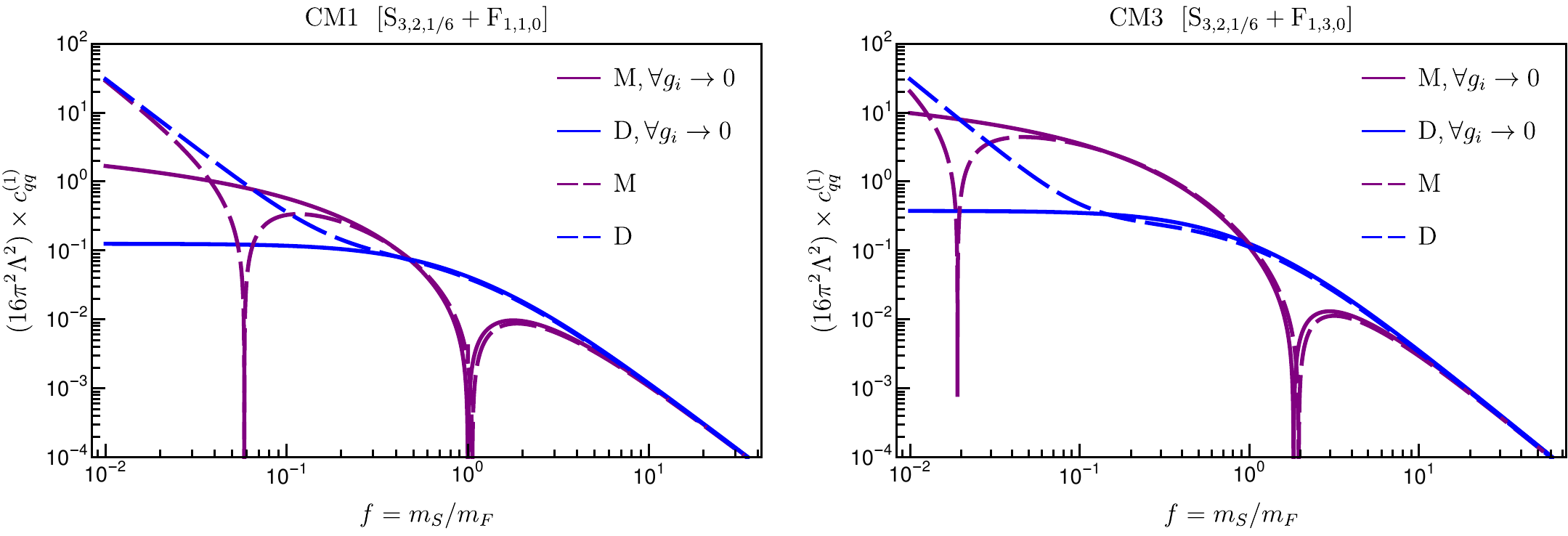}
\caption{Matching of $c_{qq}^{(1)}$ in the two coloured models CM1 (left) and
  CM3 (right) as function of $f=m_S/m_F$. The plots compare the matching with
  and without SM gauge couplings. The BSM Yukawas are assumed to be
  equal to one. Here, M and D stands for the Majorana and Dirac versions of CM1 and CM3. $\Lambda=m_F$.
}
\label{fig:OllMtchC1}
\end{center}
\end{figure}

Let us discuss the matching for CM1 and CM3 first. In the limit 
of vanishing SM couplings, CM1 generates only ${\cal O}_{qq}^{(1)}$, 
by construction, while CM3 generates both ${\cal O}_{qq}^{(1)}$ 
and ${\cal O}_{qq}^{(3)}$. In Fig.\ \ref{fig:OllMtchC1} we show 
the matching results as a function of $f=m_S/m_F$ for CM1 (left) 
and CM3 (right). The plots compare the numerical results for 
the matching for vanishing SM couplings to the full result, as 
calculated with \verb|MatchmakerEFT|. 

A few comments are in order. First of all, for model CM1 one finds
again the ``Majorana cancellation'' at $f=1$. For model CM3 the same
effect leads to a cancellation at $f\simeq 2$. We also show the
matching results assuming that $F_{1,n,0}$ are Dirac fields, to demonstrate
that this cancellation is present only for Majorana fermions. For CM1
one can see that for $f\gsim 0.2$ neglecting the SM couplings in the
matching has only a very minor effect, but for scalar masses much
smaller than the fermion mass, corrections to the matching from SM
gauge couplings lead to sizeable changes in the result. For CM3 the
results are similar for the Dirac case, but for a Majorana $F_{1,3,0}$,
neglecting SM gauge couplings, changes the matching coefficient by a
considerable factor for $f\ll 1$.  
Just for completeness, we summarise the matching results for the different models in Table \ref{tab:coefficients_CM1_CM3}.

\begin{table}[h]
\centering
\begin{tabular}{|c|c|c|c|c|}
\hline
model & CM1 & CM1-D & CM3 & CM3-D  \\
\hline
\hline
$\tilde{c}_{qq}^{(1)}$ & 0 &  $-\frac{1}{24}$ & $\frac{1}{9}$ & $-\frac{1}{8}$ \\
\hline
$\tilde{c}_{qq}^{(3)}$ & 0 & 0 & $-\frac{1}{6}$ & $-\frac{1}{12}$ \\
\hline
\end{tabular}
\caption{Coefficients $\tilde{c}_{qq}^{(1)} = (16 \pi^2 \Lambda^2) \times c_{qq}^{(1)}$ and $\tilde{c}_{qq}^{(3)} = (16 \pi^2 \Lambda^2) \times c_{qq}^{(3)}$ for $f\rightarrow 1$ for different coloured models with BSM Yukawas equal to one and vanishing SM couplings. See text for details.}
\label{tab:coefficients_CM1_CM3}
\end{table}

Model CM1 (CM3) produces only one (two) operators proportional to
$Y^4$. However, because both models contain a BSM coloured scalar,
many other 4-quark operators are generated with terms proportional to
$g_3^4$ and also $g_3^2|Y|^2$. Since $g_3$ is still large at LHC
energies, these contributions are not negligible. As an example, we
list the matching of all quark operators for model CM1 (Majorana),
neglecting $g_1$ and $g_2$, in the limit $m_S=m_F$, in Table
\ref{tab:MtchCM1}. Note that the octet operators ${\cal O}_{ud}^{(8)}$,
${\cal O}_{qu}^{(8)}$ and ${\cal O}_{qd}^{(8)}$ have the largest coefficients,
while a number of other coefficients are still zero in this
approximation. Also it is important to point out once more that
contributions proportional to $g_3^4$ are, of course, generation
diagonal. We also calculated the matching assuming Dirac fermions instead and found that only $\tilde{c}_{qq}^{(1)}$ differs by an additional term $-\frac{1}{24}|Y|^4$ and all the other eleven coefficients do not change. 

\begin{table}[t]
\begin{tabular}{|c|c|c|c|c|c|}\hline
$\mathbf{\tilde{c}_{qq}^{(1)}}$ & $\mathbf{\tilde{c}_{qq}^{(3)}}$  & $\mathbf{\tilde{c}_{uu}}$ & $\mathbf{\tilde{c}_{dd}}$  & $\mathbf{\tilde{c}_{ud}^{(1)}}$ & $\mathbf{\tilde{c}_{ud}^{(8)}}$ \\
\hline
$\frac{1}{288}g_3^2|Y|^2 - \frac{1}{720}g_3^4$ & 
$\frac{1}{96}g_3^2|Y|^2 - \frac{1}{240}g_3^4$ & $-\frac{1}{180}g_3^4$ & 
$-\frac{1}{180}g_3^4$  & $0$ & $-\frac{1}{30}g_3^4$\\ 
\hline
\hline
$\mathbf{\tilde{c}_{qu}^{(1)}}$ & $\mathbf{\tilde{c}_{qu}^{(8)}}$  & $\mathbf{\tilde{c}_{qd}^{(1)}}$ & $\mathbf{\tilde{c}_{qd}^{(8)}}$  & $\mathbf{\tilde{c}_{quqd}^{(1)}}$ & $\mathbf{\tilde{c}_{quqd}^{(8)}}$ \\
\hline
$0$ & $\frac{1}{24}g_3^2|Y|^2 - \frac{1}{30}g_3^4$ & 
$0$ & $\frac{1}{24}g_3^2|Y|^2 - \frac{1}{30}g_3^4$ & 0 & 0 \\
\hline
\end{tabular}
\caption{Matching of $\tilde{c}_{XY} = (16 \pi^2 \Lambda^2) \times c_{XY}$ in the coloured model CM1 for Majorana fermions,
keeping $g_3$ non-zero, but neglecting the smaller contributions 
from $g_1$ and $g_2$. $|Y|^2$ is the BSM Yukawa coupling and 
generation indices are suppressed for simplicity.}
\label{tab:MtchCM1}
\end{table}

\begin{figure}[t]
\begin{center}
\includegraphics[width=0.485\linewidth]{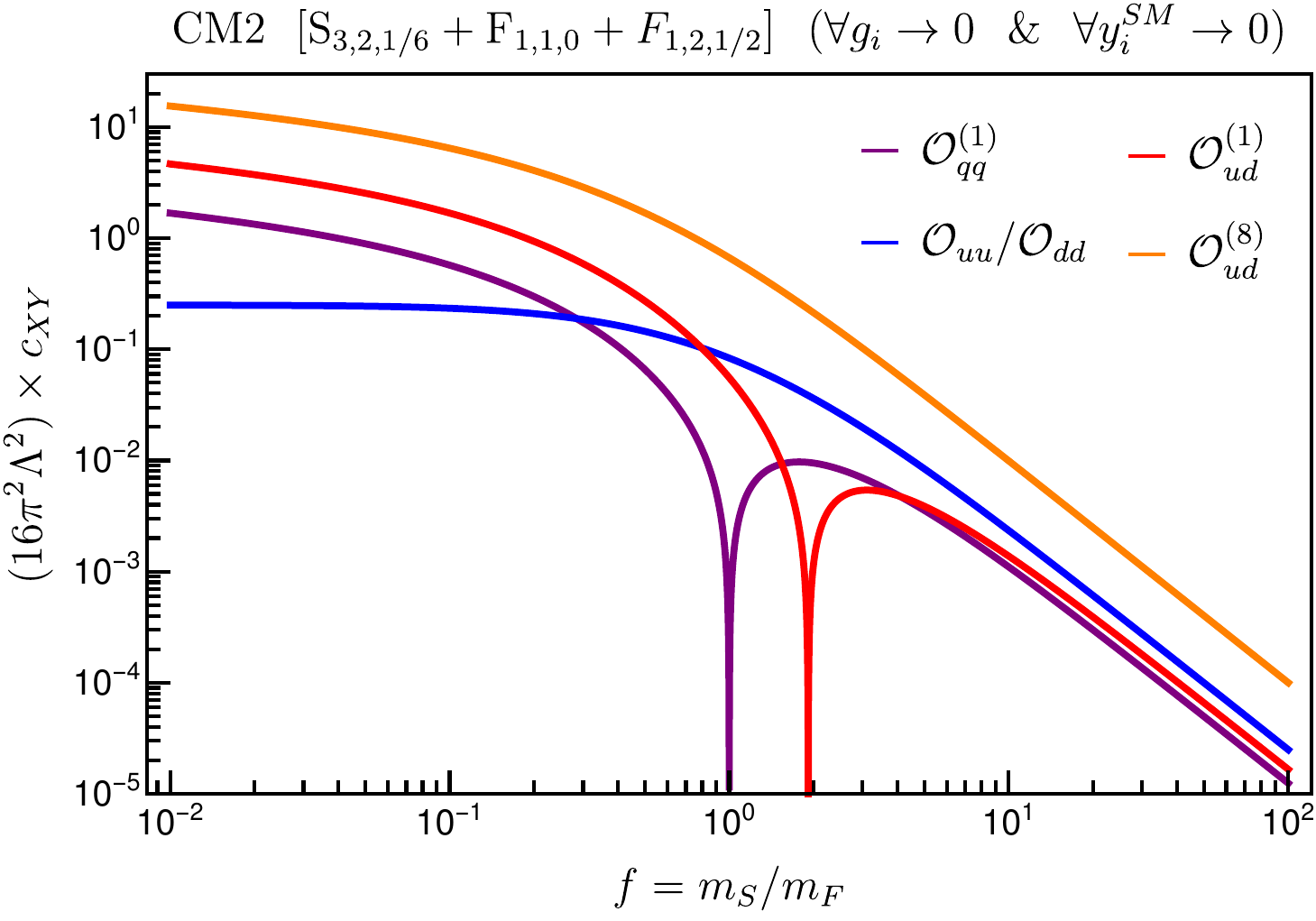}
\hfill
\includegraphics[width=0.485\linewidth]{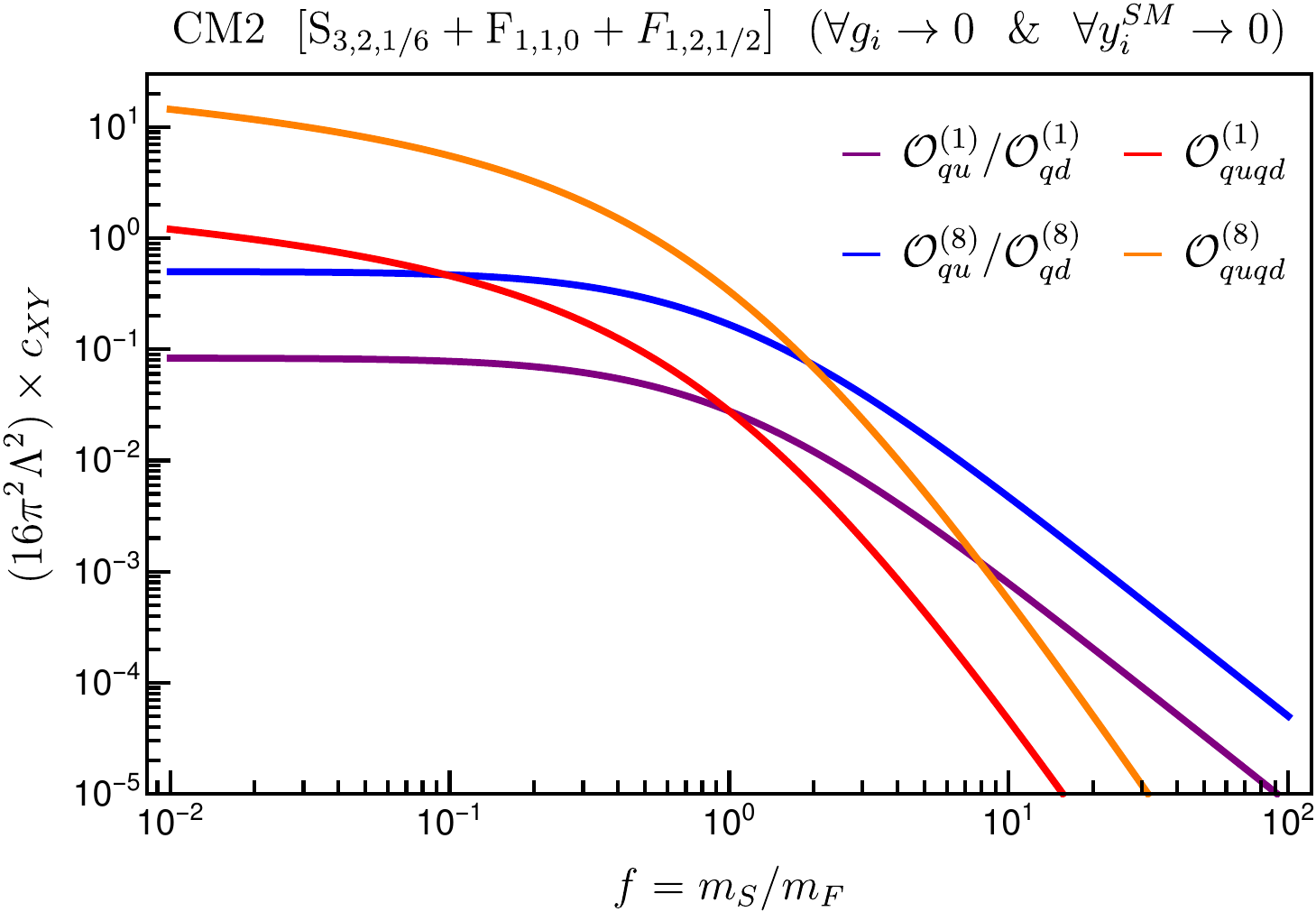}
\caption{Matching of $c_{XY}$ in the coloured model CM2 as 
function of $f=m_S/m_F$. The plots show the result in the 
limit of SM gauge couplings approaching zero. $\Lambda=m_F$.}
\label{fig:OllMtchC2}
\end{center}
\end{figure}

Model CM2 has two more Yukawa couplings, connecting up and 
down quarks to $S_{3,2,1/6}$ and $F_{1,2,1/2}$. Thus, all 4-quark 
operators are generated with terms proportional to $Y^4$ in 
this model. The numerical values of the matching coefficients 
are shown in Fig.\ \ref{fig:OllMtchC2}, again in the limit where  
SM couplings are neglected.

Imposing $m_{F_{1,2,1/2}} = m_{F_{1,1,0}} = m_F$ allows us to perform the same analysis of $c_{XY}$ as a function of $f = m_S/m_F$. As before, for $f\lsim 0.1$ 
these results are only approximate. Both ${\cal O}_{qq}^{(1)}$ and 
${\cal O}_{ud}^{(1)}$ show a cancellation at some specific value of 
$f$. We give for completeness in Table \ref{tab:coefficients_CM2_f1} the matching at $f=1$ in the 
limit where all BSM Yukawa couplings are equal to one and all SM couplings neglected.
Clearly, the pattern is very different from model CM1 and the coefficients $c_{qq}^{(1)}$, $c_{quqd}^{(1)}$, $c_{quqd}^{(8)}$ differ assuming Majorana or Dirac fermions.

\begin{table}[h]
\centering
\begin{tabular}{|c|c|c|c|c|c|c|}
\hline
$\mathbf{\tilde{c}_{XY}}$ & $\mathbf{qq^{(1)}}$ & $\mathbf{qq^{(3)}}$ & $\mathbf{uu}$ & $\mathbf{dd}$ & $\mathbf{ud^{(1)}}$ & $\mathbf{ud^{(8)}}$ \\
\hline \hline
M & 0 & 0 & $-\frac{1}{12}$ & $-\frac{1}{12}$ & $\frac{1}{18}$ & $-\frac{2}{3}$ \\
\hline
D &  $-\frac{1}{24}$ & 0 & $-\frac{1}{12}$ & $-\frac{1}{12}$ & $\frac{1}{18}$ & $-\frac{2}{3}$ \\
\hline
\hline
$\mathbf{\tilde{c}_{XY}}$ & $\mathbf{qu^{(1)}}$  & $\mathbf{qu^{(8)}}$ & $\mathbf{qd^{(1)}}$  & $\mathbf{qd^{(8)}}$ & $\mathbf{quqd^{(1)}}$ & $\mathbf{quqd^{(8)}}$ \\
\hline
M &  $-\frac{1}{36}$ & $-\frac{1}{6}$ & $-\frac{1}{36}$ & $-\frac{1}{6}$ & $-\frac{1}{36}$ & $\frac{1}{3}$ \\
\hline
D &  $-\frac{1}{36}$ & $-\frac{1}{6}$ & $-\frac{1}{36}$ & $-\frac{1}{6}$ & 0 & 0 \\
\hline
\end{tabular}
\caption{Limits for $f\rightarrow 1$ of some Wilson coefficients $\tilde{c}_{XY} = (16 \pi^2 \Lambda^2) \times c_{XY}$ in the model CM2 for both the Majorana (M) and Dirac (D) case, assuming all SM couplings zero and BSM couplings of one.}
\label{tab:coefficients_CM2_f1}
\end{table}

In contrast to CM1/CM3, the model CM2 also generates sizeable contributions to FH operators
of the form $\phi^3\psi^2$ and $\phi^2\psi^2D$. For these 
operators, CM1 and CM3 do receive contributions from diagrams with gauge 
couplings $g_{1,2}$, but none with $g_3$ and thus these operators are 
very much suppressed relative to the 4F operators in these scenarios. For CM2, 
on the other hand, the matching of FH operators 
contains terms proportional to $Y^4$ and $Y^2\lambda_3$, where 
$\lambda_3$ is the coefficient of the interaction $|S_{3,2,1/6}|^2|H|^2$, see Appendix \ref{sect:lag}. 
There is also an interesting pattern in these coefficients, depending 
on whether the fermion $F_{1,1,0}$ is assumed to be Majorana or Dirac, summarised in Table \ref{tab:coefficients_fermion_Higgs_CM2}.
Again, for simplicity, these matching coefficients have been 
written in the limit where $\forall \, Y=1$, and all heavy masses are 
equal, $m_S=m_{F_i}=\Lambda$. For better readability we multiply by an overall factor 
of $(16\pi^2 \Lambda^2)$. 

\begin{table}[h]
\centering
\begin{tabular}{|l|c|c|c|c|c|c|c|}
\hline
$\tilde{c}_{XY}$ & $Hq^{(1)}$ & $Hq^{(3)}$ & $Hu$ & $Hd$ & $Hud$ & $uH$ & $dH$ \\
\hline
\hline
Majorana & 0 & 0 & $-\frac{1}{6}$ & $\frac{1}{6}$ & $-\frac{1}{3}$ & $\frac{1}{6}(\lambda_3-2)$ & $-\frac{1}{6} (\lambda_3-2)$ \\
\hline
Dirac & $-\frac{1}{6}$ & 0 & $-\frac{1}{6}$ & $\frac{1}{6}$ & 0 & 0 & $-\frac{1}{6} (\lambda_3-1)$ \\
\hline
\end{tabular}
\caption{Wilson coefficients $\tilde{c}_{XY} = (16 \pi^2 \Lambda^2) \times c_{XY}$ for fermion-Higgs operators for $m_S = m_{F_i} = \Lambda$ in the model CM2 with all BSM Yukawas set to one and all SM couplings to zero.}
\label{tab:coefficients_fermion_Higgs_CM2}
\end{table}

In summary, we have discussed in this subsection the matching of
various example models for 4F and FH operators. Very different patterns can emerge in the different 
models, allowing in principle a model discrimination if any of 
these operators were to be observed in the future.

\begin{table}[h]
\centering
\begin{tabular}{|l|l|l|}
\hline
model & particles & BSM Yukawa couplings \\
\hline
\hline
Sc-I & $F_{1,1,0}$, $S_{1,2,1/2}$ & $Y_{\nu}$ \\
\hline
Sc-III & $F_{1,3,0}$, $S_{1,2,1/2}$ & $Y_{\Sigma}$ \\
\hline
Sc-I+ & $F_{1,1,0}$, $S_{1,2,1/2}$, $S_{1,1,1}$ & $Y_{\nu}$, $Y_{NE}$ \\
\hline
\hline
CM1 & $F_{1,1,0}$, $S_{3,2,1/6}$ & $Y_{Q1}$ \\
\hline
CM3 & $F_{1,3,0}$, $S_{3,2,1/6}$ & $Y_{Q3}$ \\
\hline
CM2 & $F_{1,1,0}$, $S_{3,2,1/6}$, $F_{1,2,1/2}$ & $Y_{Q1}, Y_{uF}, Y_{dF}$ \\
\hline
\end{tabular}
\caption{Summary of the particle content and the BSM Yukawa couplings for the classes of scotogenic and coloured models discussed here. The definition of each Yukawa is given in Appendix \ref{sect:lag}.}
\label{tab:matching_models_summary}
\end{table}


\section{Phenomenology\label{sect:pheno}}

In the previous sections we have discussed the landscape of UV models with a DM candidate and loop-suppresed 4F operators. We also identified two classes of models with distinctive SMEFT mapping, leptophilic and colored models. In this section we discuss the typical phenomenology one should expect from these scenarios as a way to illustrate the interplay between low-energy, Dark Matter and LHC phenomenology. In the last subsection, we also discuss more exotic possibilities emerging from the matching to UV theories.

In all cases, we have considered a Dark Matter candidate with hypercharge $Y=0$, a phenomenological requirement due to direct detection constraints, which point to DM candidates that do not couple to the Z-boson at tree-level. Instead, the heavy DM particle would couple to quarks via loop-suppressed interactions. 

In the leptophilic examples, we will discuss the interconnections among low-energy lepton 4F constraints, DM direct detection and relic abundance, neutrino masses, rare decays and direct collider searches. 

In the colored case, we will show how constraints from the CMB determination of the relic abundance and LHC squark-pair searches and contact interaction searches in dijets play complementary roles when exploring the parameter space. We also discuss how these new states could strengthen the electroweak phase transition to lead to strong first-order phase transition, a particularly interesting situation to explain the matter-antimatter asymmetry and production of gravitational waves in the Early Universe.

\subsection{Leptophilic models: dark matter, low-energy constraints, neutrino physics and lepton colliders}

In Sec.~\ref{subsubsect:Scot} we described the simplest models with DM candidates generating ${\cal O}_{\ell\ell}$, the so-called scotogenic model and variations. In the scotogenic scenarios, one finds two possible DM candidates, a right-handed neutrino $N_R=F_{1,1,0}$ or the lightest neutral component of a new scalar doublet $\eta=S_{1,2,1/2}$.

The phenomenology of these DM candidates has been thoroughly studied
in the literature, both for the fermionic DM
candidate~\cite{Kim:2006af,Restrepo:2013aga,Dedes:2014hga,
  Escudero:2016ksa,Hagedorn:2018spx, Jung:2020ukk,Carmona:2020uqx} and
the scalar
option~\cite{Burgess:2000yq,Cline:2013gha,Escudero:2016tzx,Escudero:2016ksa,Gross:2017dan,Avila:2019hhv,Beniwal:2020hjc,Coito:2021fgo}. In
these works, the emphasis was placed in the neutrino mass generation
and the need to accommodate the mass splittings and angles in a
neutrino fit. The neutrino fit generically leads to CLFV (``charged
lepton flavour violation'') four-fermion interactions
$c_{\ell\ell}^{ijkl}$, able to mediate rare processes like $\mu\to
e\gamma$ or $\mu\to 3 e$.  For example, the branching ratio $BR(\mu
\to e \gamma) \sim \frac{\alpha_{em}|Y_\mu|^2 |Y_e|^2 }{256 \pi
  G_F^2m_F^4}$ for $m_F \simeq
m_S$~\cite{Vicente:2014wga,Rocha-Moran:2016enp,Hagedorn:2018spx}, and
is constrained to be less than $6\times 10^{-14}$
~\cite{BasiBeneito:2020mdr}. This would lead to a limit of the order
of $\sqrt{|Y_e||Y_\mu|} \lesssim 3\times 10^{-2} (m_F/$TeV).

Moreover, the contribution to the neutrino mass from the Weinberg operator would be of the order of
\begin{equation}
m_\nu \simeq \frac{\lambda_5 \, Y^2 \, v^2}{64 \, \pi^2 \, \Lambda}
\end{equation}
see Sec.~\ref{subsubsect:Scot} for details. Assuming this mass
contribution is of the order of the eV scale, or below, then leads to
a value $\sqrt{\lambda_5} |Y| \lesssim 4\times 10^{-4} (\Lambda/\textrm{TeV})^{1/2}$ \  with $m_\nu = 1$ eV.

In summary, the neutrino fit and rare decays constraints would lead to a
scenario where the leptophilic couplings are very suppressed and the
phenomenological interplay in LHC probes would be rather difficult.
Note, however, that for $\lambda_5 \to 0$, neutrino masses would 
disappear, but the $d=6$ 4F operators remain unchanged.

In this work, though, we do not focus on describing a UV scenario with
new heavy states coupled to three generations, where dark matter, the neutrino fit,
CFLV and anomalous dipole moments could be considered at once. This is
already an active area of research, see e.g. Refs.~\cite{Li:2022chc,Hagedorn:2018spx,Avila:2019hhv,Dcruz:2022dao,Cepedello:2022xgb,Alvarez:2023dzz}.  Here, we are interested
in identifying simple models which could generate four-fermion operators and
contain DM candidates, and possibly exhibit an interesting collider signature.

\subsubsection{The fermionic DM scenario}

In the following, we will discuss the complementarity among experimental probes that the SMEFT and the UV matching is able to bring. We will simply illustrate this complementarity in a simple, yet not too trivial, leptophilic scenario. As an example, let us explore a simplified $F_{1,1,0}$ DM scenario.  

Assuming the fermion is heavy, the annihilation cross section to leptons would be given by~\cite{Herrero-Garcia:2018koq}
\begin{equation}
\langle \sigma v \rangle = \frac{m_F^2}{32 \pi (m_F^2+m_S^2)^2} \, |Y|^4 \textrm{ (Dirac, s-wave)}
\end{equation}
and 
\begin{equation}
\langle \sigma v \rangle = \frac{m_F^2 (m_F^4+m_S^4)}{8 \pi x_F (m_F^2+m_S^2)^4} \, |Y|^4 \textrm{ (Majorana, p-wave)}
\end{equation}
where $x_F=m_F/T$. The relic abundance is approximately given by $\Omega h^2 \simeq \frac{8.7\times 10^{-11} \textrm{ GeV}^{-2}}{\sqrt{g_*} \int_{x_F}^\infty \frac{d x}{x^2} \langle \sigma v \rangle}$, where $x_F \simeq 25-30$ in the TeV range, and one needs to impose $\Omega h^2 \simeq 0.1$ from the CMB constraints.

In the limit that $m_S \gtrsim m_F$ and neglecting co-annihilations~\cite{Griest:1990kh}, the relic abundance would be satisfied for Dirac fermion masses in the range
\begin{equation}
m_F \lesssim 1.1 \, |Y|^2 \textrm{ TeV (Dirac),}\label{Diracrelic}
\end{equation}
whereas for the Majorana case the value of the mass would be lower, 
\begin{equation}
m_F \lesssim 0.2 \, |Y|^2 \textrm{ TeV (Majorana).}\label{Majoranarelic}
\end{equation}
For smaller values of $m_F$, this DM candidate would only constitute a fraction of the total relic abundance.

   \begin{figure}[t!]
\begin{center}
\includegraphics[width=3.in]{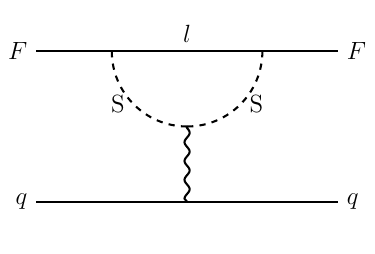}
\caption{Diagram representing a possible interaction mechanism between
  Dark Matter ($F$) and nuclei, mediated through dipole moment terms
  generated at one loop.  }
\label{fig:DDdiag}
\end{center}
\end{figure}

These fermion DM models could also lead to interesting Direct Detection (DD) phenomena, see e.g.Refs. \cite{Schmidt:2012yg,Ibarra:2016dlb}. In Ref.~\cite{Herrero-Garcia:2018koq},  bounds from DD for Majorana and Dirac $F_{1,1,0}$ were discussed. For a Dirac fermion, they found that the leading contribution to nucleon scattering was mediated by loop-induced Electron Dipole Moment, see Fig.~\ref{fig:DDdiag}. Moreover, in the EFT limit, $m_F \gtrsim$ TeV, the spin-independent cross section for scattering off nuclei is almost constant with the mass and Xenon-based experiments lead to an overall limit on the Yukawa,
\begin{equation}
Y \lesssim 10^{-2} \, \textrm{ (Dirac). }
\end{equation}
This is a strong limit on the Yukawa, which would exclude the Dirac DM option as a good  candidate for matching to an EFT.  
	
The DD limit is weakened for a Majorana DM candidate, whose cross-section decreases with the mass and it does not have dipole moments, $(\bar F \sigma_{\mu\nu} F) F^{\mu\nu}$ or $(\bar F \sigma_{\mu\nu} i \gamma^5 F) F^{\mu\nu}$.  Moreover, the DD cross-section gets even weaker as one switches on the scalar-Higgs portal coupling $\lambda_{HS}$. Indeed, in the Majorana case, the Yukawa $Y$ could be order one or higher, allowing a good EFT description while satisfying the relic abundance constraint in Eq.~\ref{Majoranarelic}.  Note that limits coming from the invisible widths of the Z boson and Higgs would be weaker than  DD limits in the heavy mass range, as discussed in~\cite{Herrero-Garcia:2018koq}. 

Co-annihilations with other states could increase the annihilation cross section. For example, $m_F$ could be very close to $m_S$ or the fermionic DM could be pseudo-Dirac~\cite{DeSimone:2010tf}.  The increase in $\langle \sigma v \rangle$ would weaken the relic abundance bounds in Eqs.~\ref{Diracrelic} and \ref{Majoranarelic} by a factor $\cal O$(1), allowing heavier DM candidates. Those semi-degeneracies could also impact the DD phenomenology, leading to e.g. inelastic DM~\cite{Tucker-Smith:2001myb}, or pointing towards displaced vertices signatures at the LHC~\cite{DeSimone:2010tf}. 

As discussed above, scenarios where DM could mediate 4F operators contributing to CLFV would lead to very stringent limits, and one could ask whether DD could be competitive with CLFV.  In Ref.~\cite{Herrero-Garcia:2018koq} the authors showed that for heavy scalars and fermions, the best sensitivity to Majorana DM would come from CLFV, if the DM mediates those. For the Dirac DM case, which we have seen already is excluded as a good EFT match, DD and CLFV are competitive sources of information.

All this discussion, where we have taken into account flavour and neutrino fit constraints, CMB relic abundance and direct detection Xenon experiments, points to an interesting  DM and 4F interplay if   the DM candidate is Majorana and does not mediate CLFV.  Let us then discuss what is the status of this option.

As discussed in Sec.~\ref{subsect:match}, the 4F operator
$c_{\ell\ell}$ is the only operator generated by the leptophilic
model Sc-I and is given by $c_{\ell\ell}= f(m_S/m_F) |Y|^4/(16 \pi^2 \Lambda^2)$. In
Fig.~\ref{fig:OllMtch2} we showed that $f(m_S/m_F) |Y|^4 \lesssim
10^{-2}$ for $m_F< m_S$ (purple-solid line in this figure), quickly
decreasing as the ratio $m_S/m_F$ increases.

Using the $\chi^2(c)$ results given in Ref.~\cite{Falkowski:2017pss} and particularising to the case where all the EFT coefficients are zero, except $c_{\ell\ell}$, we find a bound at 2-$\sigma$ for $\bar c_{\ell\ell} = c_{\ell\ell} v^2 =[-1.0, 6.1]$ $\times 10^{-3}$. This bound on the 4F operator leads to a parameter range $\frac{|Y|^4 v^2}{16 \pi^2 m^2}\simeq 10^{-1}$ for $m_S \gtrsim m_F$, which then turns into an approximate relation $|Y|^2/m \lesssim 16 $ TeV$^{-1}$. Combining both the 4F limit and the DM relic abundance limit in Eq.~\ref{Majoranarelic}, we obtain an approximate range where the fermionic DM could accommodate both the DM abundance of the Universe and the current low-energy limits on 4F operators
\begin{equation}
0.1 |Y|^2 \lesssim m_F \textrm{ (TeV)} \lesssim 0.2 |Y|^2 \, \ ,
\label{rangemF}
\end{equation}
although one should keep in mind that in this range calculation we
have made approximations which work in different directions: {\it 1.)}
We neglected co-annihilations with $S_{1,2,1/2}$, which would weaken the upper bound and allow a
higher mass range, and {\it 2.)} we assumed that $m_F$ is lower than
$m_S$, but not very far from it. Otherwise, the lower limit would be
strengthened, leaving a smaller parameter space.

\begin{figure}[t!]
\begin{center}
\includegraphics[width=3.5in]{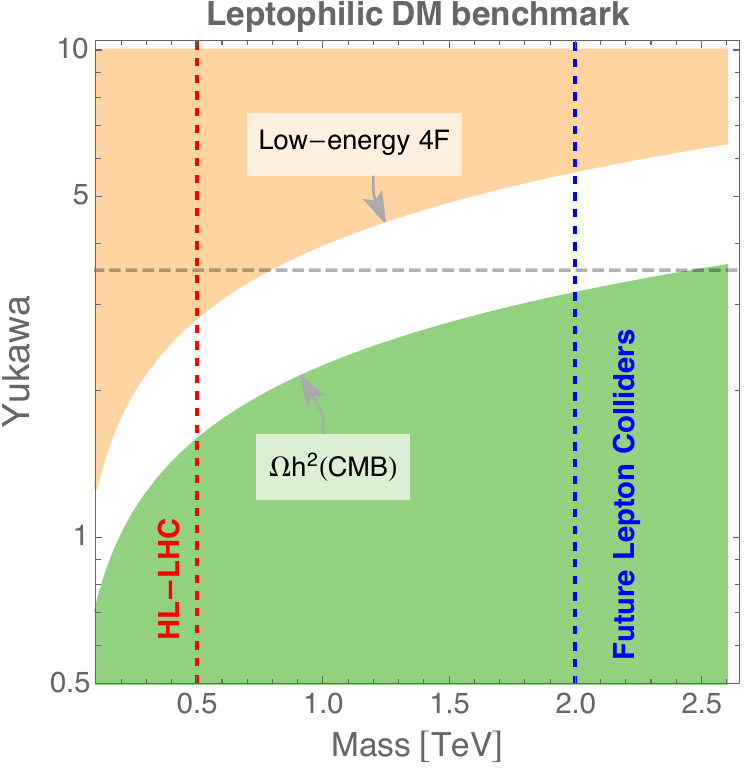}
\caption{Summary of mass vs coupling constraints in the leptophilic UV
  model with particle content $F_{1,1,0}$ and $S_{1,2,1/2}$. The
  excluded region from low-energy 4F constraints on the operator
  ${\cal O}_{\ell\ell}$ is shown in orange, whereas in green we show
  the region excluded by the condition that the DM $F_{1,1,0}$
  candidate does not overclose the Universe. The red and blue dashed
  lines correspond to prospects of 95\% CL exclusions for HL-LHC and
  future lepton colliders, respectively, from
  Ref.~\cite{Baumholzer:2019twf}. We also show with a gray-dashed line
  the non-perturbative frontier $Y=\sqrt{4\pi}$. Larger values of
    $Y$ can not lead to realistic models.}
\label{fig:leptosummary}
\end{center}
\end{figure}

In summary, a fermionic DM leptophilic scenario is a viable option,
assuming the new states couple preferentially to one lepton generation
and the DM is Majorana. This scenario could explain DM and evade 4F
constraints from low-energy precision measurements as long as $m_F
\simeq 0.2 |Y|^2$ TeV. Once these constraints are taken into account,
the allowed parameter space in mass versus Yukawa is a band, as shown
in Fig.~\ref{fig:leptosummary}. In this figure we plot the excluded
region by low-energy constraints in orange color, and in green the
region of mass and coupling which would lead to overclosing of the
Universe, also excluded.

As a side remark, we mention that in deriving the above constraints, 
we always assume that the scalar in the scotogenic does not acquire 
a vacuum expectation value, i.e. that the underlying stablizing 
symmetry is not broken spontaneously. This requires the choice $m_S^2>0$. 
One may wonder, whether points with positive $m_S^2$ are driven to 
negative $m_S^2$ under RGE running, thus ruling out sizeable parts 
of parameter space of such models as phenomenologically acceptable 
explanation for the dark matter problem \cite{Merle:2015gea}. 
However, it was shown in \cite{Alvarez:2021otp} that points with 
$m_S^2>0$ at eletro-weak scale energies never break the $Z_2$ 
at earlier times, essentially due to finite temperature effects.

Finally, in Fig.~\ref{fig:leptosummary}, we plot the prospects for
direct detection of the model at colliders, discussed in
Ref.~\cite{Baumholzer:2019twf}. At colliders, the best handle would
come from pair production of the $S_{1,2,1/2}$ through its electroweak
couplings, and the subsequent decay into $F_{1,1,0}$ and a
lepton. Whereas the prospect is that the HL-LHC would probe the mass
range around 500 GeVs (red-dashed line), future colliders could dip
into the 1.5-2 TeV range for $m_{S,F}$ (blue-dashed). The collider
probes would then cover quite a lot of the parameter space in
Eq.~\ref{rangemF} for perturbative Yukawas, $Y\lesssim \sqrt{4 \pi}$
(grey line). Note, though, that these limits could be weakened if the
splitting between the fermion and scalar is so small that the
signature to look for is two displaced vertices with missing energy
and a lepton~\cite{Cottin:2018nms,Cottin:2022nwp}, or if the femionic
DM is pseudo-Dirac~\cite{DeSimone:2010tf} instead of pure Majorana or
Dirac, which would lead again to displaced
vertices~\cite{Davoli:2017swj}.

\subsection{Coloured models: beyond-SUSY dark matter, contact interactions, squark searches and gravitational waves}

In Sec.~\ref{subsubsect:Col}, we proposed quark-specific simple extensions of the SM producing quark-specific 4F operators, i.e\ only 4F operators with quark-quark and not lepton-quark or lepton-lepton. These were
\begin{itemize}
  \item CM1: $F_{1,1,0}$  and $S_{3,2,1/6}$,
  \item CM2: CM1 + $F_{1,2,1/2}$, and
  \item CM3: $F_{1,3,0}$ and $S_{3,2,1/6}$  
\end{itemize}
If we think about these models in terms of Supersymmetry (SUSY), $F_{1,1(3),0}$ would correspond to the quantum numbers of the Bino $\tilde B$ and Wino $\tilde W$, whereas the coloured scalar doublet $S_{3,2,1/6}$ would share the quantum numbers with the squark doublet $\tilde Q_L$. 

Although these particles have the same quantum numbers as the well-known SUSY particles, their interactions are not as restricted as in SUSY, and viable dark matter scenarios with the Bino-like and Wino-like particles could be found. 

\begin{figure}[t!]
\begin{center}
\includegraphics[width=2in]{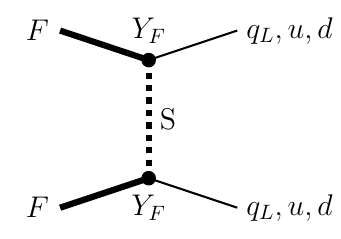}
\caption{t-channel annihilation diagram of the DM candidate $F_{1,1,0}$ into quarks.}
\label{CM1p1}
\end{center}
\end{figure}

For example, in SUSY a pure Bino scenario is disfavoured due to the smallness of its annihilation cross-section, proportional to $g_1^4$ and p-wave suppressed, which typically leads to overclosure of the Universe unless the SUSY particles are very light, in contradiction with the absence of direct observation~\cite{Arkani-Hamed:2006wnf}. But, while the Bino is necessarily coupled to SM fermions and sfermions via the SM coupling $g_1$, our DM candidate $F_{1,1,0}$ can annihilate efficiently to SM particles via a t-channel exchange of $S_{3,2,1/6}$, see Fig.~\ref{CM1p1}. In this case the cross-section is modulated by a coupling  $Y_F$: 
\begin{equation}
\label{CM1DM}
\langle \sigma v \rangle_{F_{1,1,0}}\simeq \frac{3 Y_F^4}{2 \pi \, m_S^2} \frac{r (1+r)^2}{x  (1+r)^4}
\end{equation}
where $\lambda$ here represent the coupling between $S$, $F$ and SM fermions, $r=m_F/m_S$ and $x=m_F/T$. 

Similarly, the SUSY pure Wino scenario is also disfavoured as the Wino would annihilate too efficiently into the SM via its larger $g_2$ coupling and the enhancement due to the coannihilations among the weak triplet components. This efficient annihilation of DM in the early Universe would mean that the required Wino mass to satisfy the relic abundance condition is large. On the other hand, in the  case of $F_{1,3,0}$, the coannihilation enhancement would still be present, but the overall cross-section would be modulated by a new coupling with $S$ and a SM fermion. A small coupling, smaller than $g_2$, would open up the parameter space for the Wino-like $F_{1,3,0}$ into smaller masses:
\begin{equation}
\label{CM3DM}
\langle \sigma v \rangle_{F_{1,3,0}}\simeq  \frac{3 Y_F^4}{16 \pi \, m_F^2}  \ .
\end{equation}

Those models would also induce four-quark interactions at low energies, and the specific matching is explained in Sec.~\ref{subsubsect:Col}. For these operators, the best probes come from hadron colliders. Searches for contact interactions involving four-quarks have been performed at LEP, TeVatron and now at the LHC, see e.g.~\cite{ATLAS:2017eqx,CMS:2018ucw}. 

In these experimental analyses, the typical theoretical framework to interpret the data is in terms of the following Lagrangian:
\begin{eqnarray}
\label{ }
{\cal L}_{exp} = \frac{2 \pi}{\Lambda^2} \, ( \eta_{LL} (\bar q_L \gamma_\mu q_L) (\bar q_L \gamma^\mu q_L) + 
\eta_{RR} (\bar q_R \gamma_\mu q_R) (\bar q_R \gamma^\mu q_R)+2 \eta_{RL} (\bar q_R \gamma_\mu q_R) (\bar q_L \gamma^\mu q_L)) \ . \nonumber \\ 
\end{eqnarray}
Note that, compared with the SMEFT framework: {\it 1.)} this
Lagrangian is not $SU(2)$ invariant, and {\it 2.)} there is $2\pi$
 pre-factor. The term $\eta_{LL}$ is related to the operator ${\cal
  O}_{qq}^{(1)}$, $\eta_{RR}$ to the operators ${\cal O}_{uu, dd,
  ud}^{(1)}$ and $\eta_{LR}$ to ${\cal O}_{qu, qd}^{(1)}$.

With this Lagrangian, the most up-to-date experimental limits on quark contact interactions are obtained using dijet differential distributions from the 13 TeV data. The analysis by ATLAS~\cite{ATLAS:2017eqx} is restricted to switching on only $LL$ terms and leads to the following 95\% CL limits
\begin{equation}
\label{ }
\Lambda_{LL} > 13 - 22 \textrm{ TeV, for } \eta_{LL}=\mp 1 \textrm{ (ATLAS 13 TeV 37 fb}^{-1}) \ , 
\end{equation}
whereas CMS has performed a similar analysis~\cite{CMS:2018ucw}, with comparable sensitivity. 

This experimental limit can be expressed in terms of the SMEFT operators by comparing terms $2\pi/\Lambda_{exp}^2$ with the matching results for the coefficient $c_{qq}$ in Sec.~\ref{subsubsect:Col},
\begin{equation}
\frac{2 \pi}{\Lambda_{LL}^2} = c_{qq} =   \frac{\tilde{c}_{qq}(f)}{16 \pi^2 \Lambda^2}|Y_F|^4
\end{equation}
where $f = m_S/m_F$. For $m_S \gtrsim m_F$ one finds $g(f \gtrsim 1) |Y_F|^4 \approx 10^{-2}$ as shown in Fig.~\ref{fig:OllMtchC1}. This would lead to an approximate Run2 limit 
\begin{equation}
\label{dijet}
\Lambda \gtrsim  (0.04-0.07) \, |Y_F|^2 \textrm{ TeV}
\end{equation}
Note that in Ref.~\cite{Domenech:2012ai}, the authors reinterpreted 7 TeV data in the context of 4F SMEFT. Their individual bounds were in the range $\Lambda_{eff}$= 0.8-3.5 TeV for the different four-quark operators. Also, in a more recent paper~\cite{Bordone:2021cca} a recasting of 13 TeV dijet limits was done, using the EFT approach and focused their interpretation on the flavour anomalies.
All these dijet limits are in the TeV range, but in our loop-induced models, they translate into a very weak limit on the EFT scale which jeopardizes the EFT validity. For example, if we take the Run2 limit in Eq.~\ref{dijet} and assume large values for $Y_F= \sqrt{4 \pi}$,  we  would end up with a mass limit of the order of 500-900 GeV.  But the LHC experimental dijet selection cuts are strict, typically $m_{jj}> $ 800 GeV, so even in the extreme case of large Yukawa couplings we would find that the limit on the EFT scale $\Lambda < m_{jj}^{min}=\sqrt{\hat s_{min}}$, which is in clear contradiction with an adequate EFT expansion.

\begin{figure}[t!]
\begin{center}
\includegraphics[width=2in]{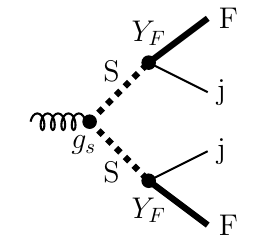}
\caption{Pair-production of the scalar $S_{3,2,1/6}$ leading to jets and missing energy from the DM $F_{1,1,0}$.}
\label{CM1p2}
\end{center}
\end{figure}

On the other hand, hadron colliders  are sensitive to direct production of the coloured particles, through processes
\begin{equation}
\label{ }
p \, p \to S_{3,2,1/6} \, S^*_{3,2,1/6} \to F_{1,1,0} \, F_{1,1,0}\, + 2 \, q \to  MET + 2 \, j \ ,
\end{equation}
see Fig.~\ref{CM1p2}. Note that this decay could go through a final state of displaced jets if $\Delta m_{FS}=m_S-m_F\ll m_S$, rendering $S$ long-lived. 

\begin{figure}[ht!]
\begin{center}
\includegraphics[width=3.5in]{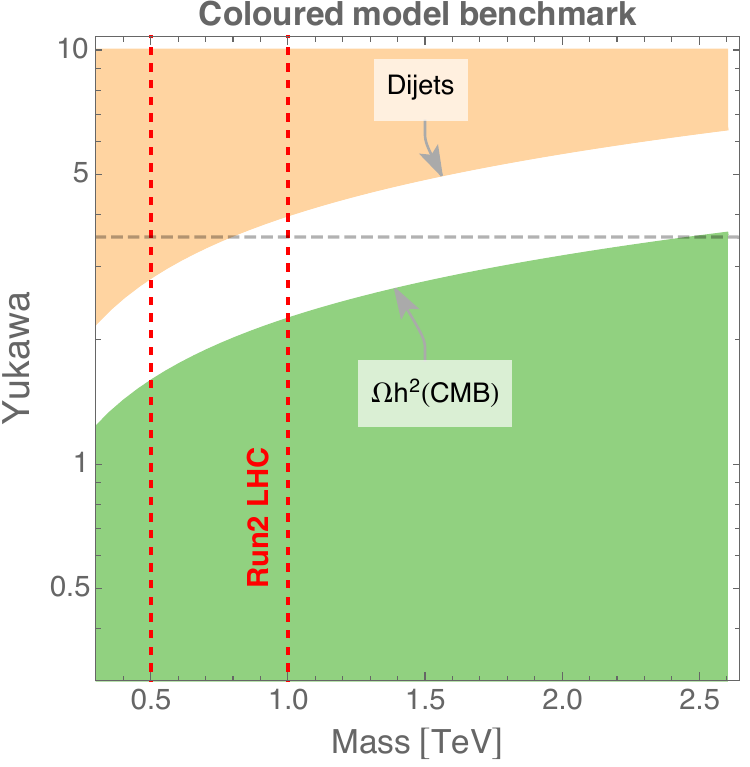}
\caption{Summary of mass vs coupling constraints in the CM1 UV model with particle content $F_{1,1,0}$ and $S_{3,2,1/6}$. The excluded region from the Run2 dijet analysis from the  ATLAS~\cite{ATLAS:2017eqx} and CMS~\cite{CMS:2018ucw} contact interaction interpretation is shown in orange, whereas in green we show the region excluded by the condition that the DM $F_{1,1,0}$ candidate does not overclose the Universe. The red  dashed lines corresponds to the range of current 95\% CL exclusions from a search for pairs of squarks~\cite{ATLAS:2020syg}. We also show with a grey-dashed line the non-perturbative frontier $Y=\sqrt{4\pi}$.    }
\label{summaryquark}
\end{center}
\end{figure}

For prompt $S$ decays, this search is exactly the same as the traditional 1st-2nd generation squark pair production leading to MET+jets. ATLAS has performed a search in this final state using the full Run2 data~\cite{ATLAS:2020syg}. We can simply translate the limits for the benchmark "{\it 1 non-degenerate $\tilde q$}" into our scenario,
\begin{equation}
\label{lim1}
m_S > 1 \textrm{ TeV, for } m_F < 400 \textrm{ GeV} \ ,
\end{equation}
a limit that quickly weakens as $m_F$ gets closer to $m_S$:
\begin{equation}
\label{lim2}
m_S > 0.5 \textrm{ TeV, for } m_F \lesssim m_S   \ ,
\end{equation}
where we still have enough kinematic space to produce a prompt decay.

Let us now show the impact of these constraints (DM relic abundance, effective four-quark operators and direct searches) in the parameter space of mass versus coupling with the SM fermions. As an example, we focus on the CM1 model with singlet Bino-like DM. The annihilation cross-section given in Eq.~\ref{CM1DM}, leads to an allowed parameter space
\begin{equation}
\label{ }
m_S \lesssim 1.7 |Y_F|^2 \textrm{ TeV (CM1).}
\end{equation}
In Fig.~\ref{summaryquark} we show the allowed region in green. We also show the current limit from dijet searches for contact interactions, using Eq.~\ref{dijet} and choosing a nominal value of 5 TeV in the range. 

The direct search limits on pair production of $S$ particles in the final state MET+ jets discussed in  Eqs.~\ref{lim1} and ~\ref{lim2} are shown as dashed red lines, and we also note the value $Y_F=\sqrt{4 \pi}$ with a dashed-grey line.

With this figure we see that the current Run2 data, including dijet searches and squark pair production, has probed a substantial portion of the region allowed by the relic abundance and perturbativity. If the squark searches reached the 2 TeV range, this scenario would be completely covered.

\subsubsection{Baryogensis and Gravitational waves}

The new scalar $S$, which enables the strong production at the LHC, could also enhance the Electroweak Phase Transition via its coupling to the Higgs $\lambda_3$, see Lagrangian \ref{eq:LagCM1}.
 
 Although the colored $S$ should not acquire a vev \cite{Casas:1995pd}, at one-loop this coupling modifies the Higgs potential.  This modification could open the window for a strong 1st order phase transition, one of the requirements for successful baryogenesis, which should be complemented by new CP violating sources, and also offers the opportunity for production of Gravitational Waves. 

Indeed, one-loop diagrams with pairs of $S$ modify the steepness of the Higgs potential via its $c_H$ term,
\begin{equation}
\frac{c_H}{\Lambda^2} \, [\phi^\dagger \phi]^3 \ .
\end{equation}
After electroweak symmetry breaking, this dimension-six term will induce a temperature correction that could increase the barrier between the true and false vacua. This barrier could change the typical SM cross-over potential into a potential that at the electroweak scale leads to a strong 1st order phase transition. 

Neglecting SM couplings, we find that the contribution of $S$ to the SMEFT coefficient is the same for  CM1, CM2 and CM3,
\begin{equation}
\frac{c_H}{\Lambda^2}= -\lambda_3 \left(\frac{\lambda_3}{4 \pi m_{S}}\right)^2,
\end{equation}
The condition that the Higgs potential should be bounded by below leads to $\lambda_3 < 0$, to  ensure that the $h^6$ term does not de-stabilize the potential at large field values. Moreover, there are upper and lower bounds on this term by imposing that the phase transition completes successfully  and that it is  strong-enough first-order, respectively~\cite{Gorbahn:2015gxa}  
\begin{equation}
0.105 < c_H\, \frac{v^2}{\Lambda^2} < 0.211 \ .
\end{equation}
This range can be written as 
\begin{equation}
m_{S}(TeV) \in [0.04,0.06] \, \lambda_3^{3/2} \ .
\end{equation}
Taking into account the current bounds, $m_{S} \gtrsim$ 0.5-1 TeV, the coupling $\lambda_3$ would have to be larger than 1 to accommodate a strong 1st order phase transition.  Even the limiting case of very large quartic coupling, $\lambda_3 \simeq 4 \pi$,  would lead to a range $m_{S} \in [1.8, 2.7]$ TeV, within the reach of future LHC searches for squarks. Therefore, we find  that the LHC direct searches will be able to explore the scalar parameter space relevant to baryogenesis and gravitational wave production in these colored scenarios.

\subsection{New particles with very exotic quantum numbers}

So far in this section we have discussed the phenomenology of a subset of paradigmatic UV completions arising from our diagrammatic analysis. We have identified a few benchmark scenarios where we could explicitly show the interplay between direct and indirect probes. 

All the benchmarks we have studied (leptophilic and coloured) contains new particles with typical quantum numbers in theories of Beyond the Standard Model, e.g. the fermionic DM particle $F_{110}$ has the same SM quantum numbers as the Bino.

Yet the models we found are far richer than the benchmarks we have studied as can be seen in Fig.~\ref{fig:overlap_matrix_singlet}, where we show the number of models with singlet DM and the structure of SMEFT operators they generate~\footnote{We have added an ancillary file to the paper with a list of models  with their matter content and the mapping to which operators they contribute to.}.

In particular, we find quite exotic new particles such as scalar or fermion color octets and weak doublet/triplet, $(S/F)_{82Y}$ and $(S/F)_{83Y}$, with different values of hypercharge $Y$. 

For example, UV completions leading to ${\cal O}_{qq}$ operators contain:
\begin{equation}
\textrm{Exotic 1 : } 	F_{{\bf 8}30} = \left(F^+_{{\bf 8}}, F^0_{{\bf 8}}, F^-_{{\bf 8}} \right), \, S_{{\bf 3} \, 2 \, 1/6}=\left( S^{+2/3}_{{\bf 3}},\, S^{-2/3}_{{\bf 3}}\right), \,  F_{110}	\ ,
\end{equation}
and in the list one also finds the same model, except with all the spins flipped, namely
\begin{equation}
\textrm{Exotic 2 : } 	S_{{\bf 8}30} = \left(S^+_{{\bf 8}}, S^0_{{\bf 8}}, S^-_{{\bf 8}} \right), \, F_{{\bf 3} \, 2 \, 1/6}=\left( F^{+2/3}_{{\bf 3}},\, F^{-2/3}_{{\bf 3}}\right), \,  S_{110}	\ .
\end{equation}
The color octects would be copiously pair produced through their strong coupling, in the same way gluinos or sgluons would do, see e.g. Ref.~\cite{Gerbush:2007fe}. Yet our UV exotic 1 and 2 models predict that, instead of one neutrally charged gluino/sgluon, one should expect a triplet of charged and neutral components, $F^{\pm,0}_{\bf 8}$ or  $S^{\pm,0}_{\bf 8}$. Exotic 1 and 2 would lead to a new type of phenomenology, with enlarged production cross section respect to the gluino or sgluon benchmarks, and opportunities to use the weak charge of the new states to tag on final states like leptonic $W$ or $Z$ decays. 

The study of the phenomenology of such exotic objects is beyond the scope of this paper, but one can identify sets of interesting channels such as  high-multiplicity events
\begin{equation}
p \, p \to F^{\pm}_{\bf 8} \, F^{\mp}_{\bf 8} \to n_\ell \textrm { leptons } + n_j \textrm { jets } + \textrm{ MET,}
\end{equation}
where $n_j\geqslant 4$ and $n_{\ell}$ would depend on whether there is enough kinematic range for a sizeable branching ratio to a $W$ or $Z$ decay, e.g. through $F^+_{\bf 8} \to W^+ \,  F^0_{\bf 8}$. These busy events, similar to long decay chains in Supersymmetry or to Black Holes~\cite{CMS:2017boz}, should be quite accessible at the LHC, as they exhibit a large production cross section, substantial missing energy, and many high-$p_T$ jets and leptons which can be used to reconstruct resonances.


\section{Conclusions}
In this work we focused on the identification of UV models that contribute to $d=6$ SMEFT 4F operators and contain a viable cold Dark Matter (DM) candidate. In particular, we have been interested in models that allow for an interplay between constraints from 4F operators and the DM relic abundance.

We classified the UV models for 4F operators with DM candidates using a diagrammatic approach. The method is similar to the approach presented in \cite{Cepedello:2022pyx}, a study where  all the new states were assumed to decay to SM particles (exit particles). Here we extended this work to allow for a Dark Matter candidate, and classify the new sets of scenarios.

Since the stable  DM candidates considered here have to be odd under a stabilising symmetry, the new states couple only in pairs to SM particles and hence all contributions to 4F operators are by construction loop-induced and a direct LHC search may be feasible. 

Also, in this paper we focused on box-diagram topologies for 4F operators. Topologies leading to portals or propagator corrections have not been considered here as they are more difficult to constrain by limits on 4F operators or provide only single-field extensions to the SM. 

Although the classification method is general, some assumptions for the box diagrams were made:
\begin{itemize}
\item We considered only models with BSM scalars and fermions, but no vectors.
\item We limited ourselves to dimension-6 operators. Higher-dimensional operators might be interesting in particular scenarios in which dimension-6 operators do not contribute to the process of interest.
\item By construction, all models contain a DM candidate. In combination with the analysis for models with exits in~\cite{Cepedello:2022pyx}, it covers all scenarios which do not feature electrically charged stable particles in the loop.
\item We neglect SM Yukawa couplings in the matching from the UV models to the SMEFT operators. Assuming natural BSM Yukawa couplings, the SM gauge couplings are also negligible in most cases.
\end{itemize}

Phenomenologically consistent DM candidates must be colour singlets and electrically neutral. Further requiring that they are	 compatible with the observed DM relic abundance and constraints from direct detection leaves us with a finite number of candidates. In this work we focused on deriving the phenomenology of scalars and fermions with hypercharge $Y=0$, i.e.\ $(S/F)_{1,n,0}$ with $SU(2)$ multiplets $n=1,3,5,...$, and on the inert doublet $S_{1,2,1/2}$.

Next, we classified the 4F operators as lepton-specific, quark-specific and mixed operators.
For the mentioned choices of DM candidates we studied the overlap between models for different 4F operators and found that many models contribute only to either lepton- or quark-specific operators. This finding differs from the results for models with exit particles as the stabilising symmetry for DM candidate prohibits SM particles appearing in the loop diagrams. 
Hence many models can be classified as lepton-specific or quark-specific models. 

Lepton-specific models receive strong constraints from low-energy limits while quark-specific models are of particular interest at hadron colliders like the LHC. Moreover, generally speaking, the sparseness of the model overlap between different operators would allow for a good model discrimination if some non-zero Wilson coefficients were experimentally established.

We then presented explicit examples for the matching for both classes: generalized versions of scotogenic models in the leptophilic case, and coloured models for the quark-specific case. We then studied the dependence of the Wilson coefficients on the mass ratio of the BSM scalars to fermions, on the Majorana or Dirac nature of the fermions and on the SM gauge couplings. 

In the lepton-specific scenario, we found a very interesting interconnection among low-energy lepton 4F constraints, the relic abundance constraint and the direct collider searches, both at the LHC and future lepton colliders. 

In the colored case, the constraints from the CMB determination of the relic abundance and LHC squark-pair searches and contact interaction searches in dijets played complementary roles when exploring the parameter space. In this case, direct searches for the colored states were more sensitive than LHC SMEFT interpretations of dijet final states. Morover, we found that these SUSY-like direct searches in the HL-LHC should be able to cover all  the region of parameter space where new states could strengthen the electroweak phase transition, with the potential to explain the matter-antimatter asymmetry and production of gravitational waves in the Early Universe.

Finally, we discussed more exotic models which arise from our analysis. In particular, we found scenarios with color-octect weak-triplet quantum numbers, which are not considered in the existing searches but should lead to very interesting phenomenology: strong production, missing energy and many hard leptons and jets.  

Attached to this paper submission, the reader can find the model files and matching results using MatchMakerEFT \cite{Carmona:2021xtq}  for the leptophillic and colored models discussed here.

\section*{Acknowledgements}

We would like to acknowledge discussions with A. Vicente, J. Herrero. We
would also like to thank Jose Santiago for help with the tool
MatchMakerEFT~\cite{Carmona:2021xtq}.  F.E. is supported by the
Generalitat Valenciana with the grant GRISOLIAP/2020/145. The research
of V.S. is supported by the Generalitat Valenciana PROMETEO/2021/083 and
the Ministerio de Ciencia e Innovacion
PID2020-113644GB-I00. M.H. acknowledges support by grants
PID2020-113775GB-I00 (AEI/10.13039/ 501100011033) and CIPROM/2021/054
(Generalitat Valenciana).  R.C. is supported by the Alexander von
Humboldt Foundation Fellowship.

\appendix

\section{Lagrangians for example models\label{sect:lag}}

In section \ref{subsect:match} we have discussed the complete 
matching for a few specific example models. For completeness and 
for fixing the notation, in this appendix we provide the Lagrangian 
terms for these models. 

We start with the scotogenic model. This model adds two new particles
to the SM content, one scalar $\eta=S_{1,2,1/2}$ and one (to three)
singlet fermion(s) $F=F_{1,1,0}$.\footnote{This particle has the same
  quantum number as a right-handed neutrino. So in the literature one
  can find notations for this field as $\nu_R$ or also $N$ and $N_R$.}
For fitting neutrino data, one needs at least two copies of
fermions. Since we will not repeat neutrino fits, we will write the
Lagrangian for only one copy of $F$, extending to more generations is
straightforward.  The new pieces of the Lagrangian, beyond the SM
terms, can be written as:
\begin{equation}\label{eq:LagSc}
{\cal L}^{\rm Sc} = \frac{1}{2}m_F \overline{F^c}F+ Y_{\nu}{\bar F}L\eta 
               + {\cal L}^{\rm Sc}_{pot},
\end{equation}
with
\begin{eqnarray}\label{eq:LagSc2}
{\cal L}^{\rm Sc}_{pot} & = & m_S^2 |\eta|^2 + \lambda_2 |\eta|^4 
                       + \lambda_3 |\eta|^2|H|^2 
                       + \lambda_4 |\eta^{\dagger}H|^2 
                       + \frac{1}{2}(\lambda_5 (\eta^{\dagger}H)^2 + {\rm h.c.})
\end{eqnarray}
For $\lambda_5\equiv 0$, this setup conserves lepton number and the 
connection with neutrino masses is lost. 

We have discussed also several variants of this basic idea. First of 
all, the model as defined above should be better called ``scotogenic 
type-I'' model. One can simply replace $F_{1,1,0}$ by $F_{1,3,0}$, 
to arrive at the scotogenic model type-III. The structure of the 
terms in eq. (\ref{eq:LagSc2}) remain the same, but in the main text 
we call the Yukawa coupling $Y_{\Sigma}$, following notation from 
neutrino physics, were $F_{1,3,0}$ is usually denoted as $\Sigma$.

To arrive at the Dirac versions of these models, we have to actually
introduce two independent Weyl spinors, $F_R$ and $F_L$. This implies
that the mass term becomes vector like ($\overline{F^c}F \to 
\overline{F_L}F_R$) and if one assigns $F_R$ and $F_L$ the same lepton 
number, no contribution to the Weinberg operator will be generated 
anymore and neutrino masses are zero.

Finally, we introduced another variant, called Sc-I$^+$ in the main 
text. This variant takes the original model and adds a second scalar, 
$S_1=S(1,1,1)$. The new Lagrangian terms are
\begin{eqnarray}\label{eq:LagScPlus}
{\cal L}^{\rm Sc^+} & = &  Y_{NE}\overline{F}eS_1 
                   + m_{S_1}^2 |S_1|^2 + \lambda_6 |S_1|^4 
                   +  \lambda_7 |S_1|^2 |\eta|^2 + \lambda_8 |S_1|^2 |H|^2
\\ \nonumber
                 & + & (\mu_S S_1^+ H \eta + {\rm h.c.}).
\end{eqnarray}

Subsection \ref{subsubsect:Col} discusses several model variants 
with a coloured scalar $S_Q = S(3,2,1/6)$ and a fermion, either 
$F=F(1,1,0)$ or $F=F(1,3,0)$. The model is reminiscent of supersymmetry, 
since $S_Q$, $F(1,1,0)$ and $F(1,3,0)$ have the same quantum numbers 
as the scalar quark, the bino and the wino. The 
lagrangian can be written as 
\begin{equation}\label{eq:LagCM1}
{\cal L}^{\rm CM} = \frac{1}{2}m_F \overline{F^c}F+ Y_{Q_i}{\bar Q}F S_Q 
               + m_{Q}^2 |S_Q|^2 + \lambda_2  |S_Q|^4 
               + \lambda_3  |S_Q|^2 |H|^2  
\end{equation}
Similar to the case of the scotogenic model, discussed above, $F$ can
stand for either the singlet ($Y_{Q_1}$) or the triplet ($Y_{Q_3}$)
and one can create easily models with either Majorana or Dirac
fermions.  Note that in supersymmetry the term proportional to $Y_F$
is fixed to be a gauge coupling (either $g_1$ or $g_2$) but we treat
$Y_F$ as a free parameter.

Finally, we introduced a variant, called CM2 in the main text, 
that adds a second fermion $F_2=F(1,2,1/2)$ to the model variant 
CM1 ($S_Q = S(3,2,1/6)$ + $F=F(1,1,0)$). New terms in the 
Lagrangian are:
\begin{equation}\label{eq:LagCM2}
{\cal L}^{\rm CM2} = m_{F_2} \overline{F_2}F_2+ Y_{HH}{\bar F}F_ 2 H^{\dagger}
                 + Y_{uF}\overline{u_R}F_2
                 + Y_{dF}\overline{F_2^c}d_R S_Q^{\dagger}. 
\end{equation}
Note that in defining all Lagrangians we have assumed the existence 
of a $Z_2$ symmetry under which the new particles are odd. Thus, 
only terms quadratic (or quartic for scalars) in the new fields 
survive.

\bibliographystyle{JHEP}
\bibliography{DM_boxes}

\end{document}